\documentclass[aps,prd,reprint,superscriptaddress,longbibliography]{revtex4-2}

\usepackage{graphicx}
\usepackage{amsmath}
\usepackage{amssymb}
\usepackage{bm}
\usepackage{hyperref}
\usepackage{xcolor}
\usepackage{upgreek}
\usepackage[normalem]{ulem}
\usepackage{tabularx}
\usepackage[mathlines]{lineno}

\definecolor{bred}{rgb}{0.8, 0.0, 0.0}
\definecolor{pblue}{rgb}{0.2, 0.2, 0.6}
\definecolor{ao}{rgb}{0.0, 0.5, 0.0}
\definecolor{carmine}{rgb}{0.59, 0.0, 0.09}

\newcommand{\CEvNS}{CE$\nu$NS}
\newcommand{\conusplus}{CONUS\texttt{+}}

\begin{document}

\title{Analysis of \conusplus~run 2: Detector performance, background model and \CEvNS~measurement}

\newcommand{\MPIK}{\affiliation{Max-Planck-Institut f\"ur Kernphysik, Saupfercheckweg 1, 69117 Heidelberg, Germany}}
\newcommand{\KIT}{\affiliation{Institut f\"ur Astroteilchenphysik, Karlsruher Institut f\"ur Technologie (KIT), Hermann-von-Helmholtz-Platz 1, 76344 Eggenstein-Leopoldshafen, Germany}}
\newcommand{\KKL}{\affiliation{Kernkraftwerk Leibstadt AG, 5325 Leibstadt, Switzerland }}
\newcommand{\MBI}{\altaffiliation[Present Address: ]{Marietta-Blau-Institut f\"ur Teilchenphysik der \"OAW, Dominikanerbastei 16, 1010 Wien, Austria}}
\newcommand{\Mirion}{\altaffiliation[Present Address: ]{Mirion Technologies GmbH, Stahlstra\ss e 42-44, 65428 R\"usselsheim, Germany}}
\newcommand{\PSI}{\altaffiliation[Present Address: ]{Paul Scherrer Institut, Forschungsstrasse 111, 5232 Villigen, Switzerland}}
\newcommand{\ESGmail}{\altaffiliation[Contact author: ]{\protect\href{mailto:esanchez@mpi-hd.mpg.de}{esanchez@mpi-hd.mpg.de}}}

\author{N.~Ackermann}\MPIK
\author{H.~Bonet}\MPIK
\author{C.~Buck}\MPIK
\author{J.~Hakenm\"uller}\MBI\MPIK
\author{G.~Heusser}\MPIK
\author{M.~Lindner}\MPIK
\author{W.~Maneschg}\Mirion\MPIK 
\author{M.~Meier}\MPIK
\author{S.~Mertens}\MPIK
\author{D.~Piani}\MPIK
\author{M.~Rank} \KKL
\author{T.~Rink} \MPIK \KIT
\author{E.~S\'{a}nchez Garc\'{i}a}\ESGmail\MPIK
\author{Y.~Shi}\MPIK
\author{L.~Sch\"onfeld}\MPIK
\author{I.~Stalder}\KKL
\author{H.~Strecker}\MPIK
\author{J.~Wendel}\MPIK
\author{J.~Woenckhaus}\PSI\KKL

\collaboration{CONUS Collaboration}
\email{conus.eb@mpi-hd.mpg.de}

\date{\today}

\begin{abstract}
Coherent elastic neutrino-nucleus scattering (\CEvNS) provides a uniquely sensitive probe of neutrino properties, but its detection requires pushing energy thresholds to the sub-keV scale while controlling backgrounds. The \conusplus~experiment, located at the Leibstadt nuclear power plant, reported the first observation of \CEvNS~using reactor antineutrinos in 2025. Following a detector upgrade that replaced three of the original 1~kg high-purity germanium detectors with larger 2.42~kg diodes, this paper presents the analysis of the \conusplus~run 2 dataset, including the detector characterization, the background model, the signal prediction and expected sensitivity, and the extraction of the \CEvNS~signal. The new detectors reach a background level per unit mass reduced by approximately a factor of two relative to the original detectors, while reaching energy thresholds between 140 and 170~eV$_{ee}$. With an exposure of 1296~kg$\cdot$days reactor-on and 86~kg$\cdot$days reactor-off, a blind likelihood analysis fitting directly the number of signal counts yields $N_s = 1565 \pm 155$ CE$\nu$NS events and a relative precision of 9.9\%. The result is in agreement with the Standard Model expectation of $1701 \pm 198$ events, with $N_s/N_{\mathrm{SM}} = 0.92 \pm 0.14$.
\end{abstract}

\maketitle

\section{\label{sec:intro}Introduction}

Among the interactions available to probe neutrino properties, coherent elastic neutrino-nucleus scattering (\CEvNS) stands out for its unusually large cross section. Since the neutrino couples coherently to the nucleus as a whole rather than to individual nucleons, it approximately scales with the square of the neutron number~\cite{Freedman:1973yd, Kopeliovich:1974mv}. This enhancement makes the process accessible to compact, kilogram-scale detectors rather than the ton-scale targets typically required for other neutrino interaction channels. The trade-off is that the signals produced are small, as coherence is preserved only when the momentum transfer is small compared to the inverse nuclear radius. A further complication arises from nuclear quenching, whereby only a fraction of the recoil energy is converted into the ionization signal actually measured by the detector, expressed in units of eV$_{ee}$ (electron equivalent energy). Therefore, \CEvNS~detection requires pushing the energy thresholds as low as possible while ensuring accurate characterization and effective suppression of background contributions. In \conusplus, the \CEvNS~region of interest (ROI) extends from the energy threshold of each detector, between 140 and 170~eV$_{ee}$ (Sec.~IV~B), up to 800~eV$_{ee}$.

\CEvNS~was first observed by the COHERENT Collaboration at a spallation neutron source on CsI~\cite{Coherent:2017},  and has since been measured there also on
argon~\cite{COHERENT:2020iec} and germanium~\cite{COHERENT:2026yje}. The interaction channel has also been detected from $^{8}$B solar neutrinos on xenon by dark matter experiments~\cite{XENON:2026smt,PandaX:2024muv,LZ:2025igz}. Reactor antineutrinos, with energies below $\sim$10~MeV, provide a third, fully-coherent source, and are being pursued by a broad range of experiments worldwide~\cite{CONNIE:2024pwt, MINER:2025cevns, Colaresi:2022obx, NEON:2022hbk, NUCLEUS:2026pnv, nGeN:2025hsd, Ackermann:2024kxo, Akimov:2024lnl, Bailly-Salins:2026inm, Dai:2025kai, Xie:2025frh, TEXONO:2026eyr}. Among these, \conusplus~reported the first observation of reactor \CEvNS~in 2025, at a significance of 3.7~$\sigma$ using an exposure of 327~kg$\cdot$days reactor-on and 60~kg$\cdot$days reactor-off~\cite{Ackermann:2025obx}. That result was limited by the statistical uncertainty on the extracted signal.

Building on this first observation, \conusplus~underwent a detector upgrade in November 2024, replacing three of the original 1~kg detectors with new, larger 2.42~kg diodes. These new detectors also achieved an improved background level per unit mass, mostly due to a more favorable mass-to-surface ratio of the larger crystals. In this paper, we present this second phase of data acquisition, \conusplus~run 2, collected with the upgraded detector array. We describe the experimental setup and detector performance, the data-taking conditions and event selection, the background model, and the resulting \CEvNS~signal prediction and sensitivity for the accumulated run 2 exposure of 1296~kg$\cdot$days. Finally, the extraction of the \CEvNS~signal and comparison with the Standard Model (SM) prediction is presented. 

\section{\label{sec:exp}Experimental set-up}

\conusplus~is located inside the reactor building of the Leibstadt nuclear power plant (Switzerland), at a distance of 20.7~m from the 3.6~GW (thermal power) reactor core, providing an antineutrino flux of $10^{13}~\mathrm{cm^{-2}\,s^{-1}}$ at the detector position. The experimental site provides an average overburden of 7.4~m of water equivalent (m w.e.), which increases by a further 0.25~m~w.e. during reactor-off periods, when the
reactor drywell head is placed over the experimental room. Prior to detector installation, the background environment at the site was characterized in detail and reactor correlated backgrounds found to have negligible impact on the \CEvNS~ROI, as described in \cite{CONUS:2024vyx}.

\begin{figure}[htb]
    \centering
    \includegraphics[width=\columnwidth]{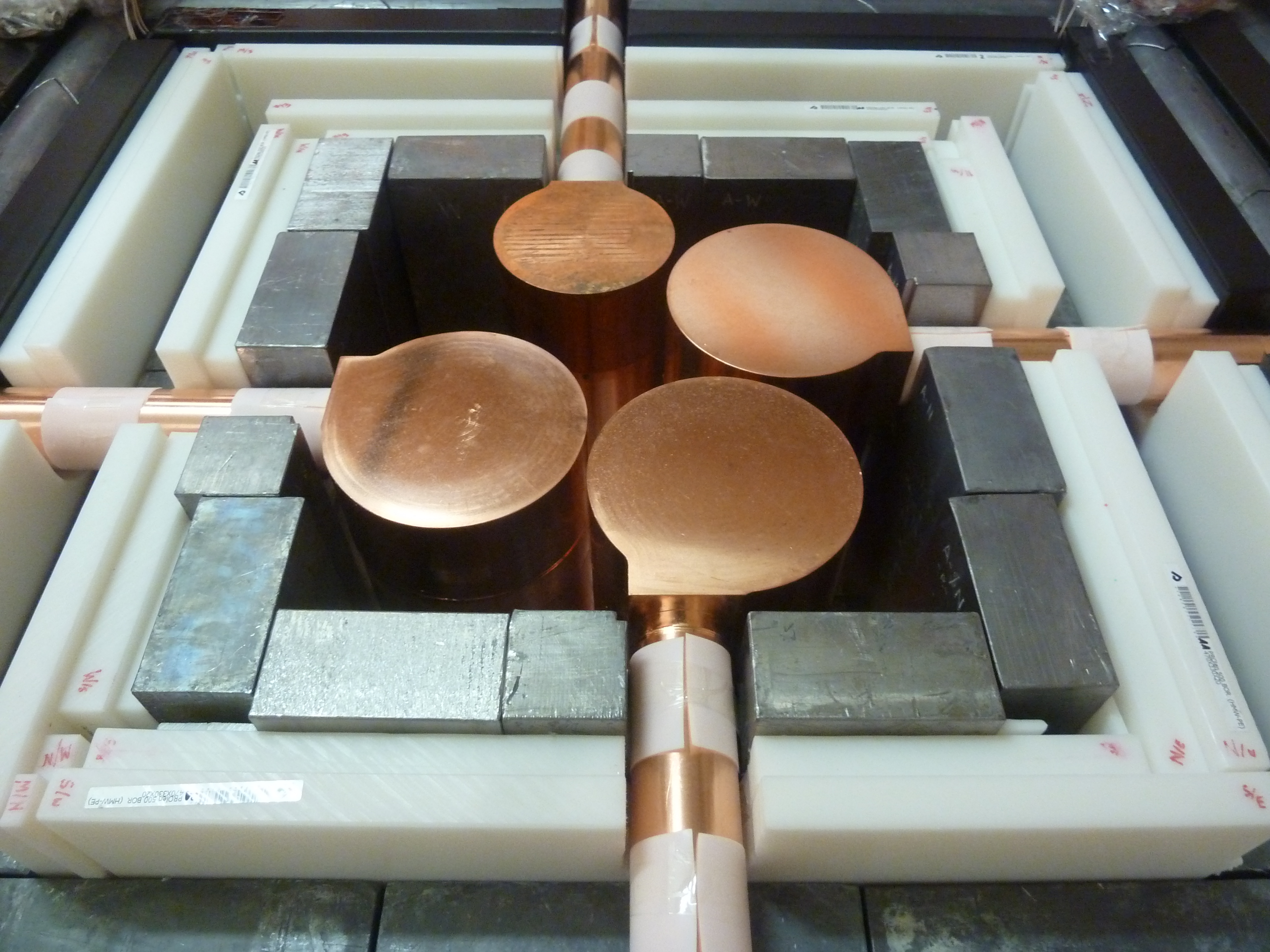}
    \caption{Top view of the open \conusplus~shield in the run 2 configuration. The four HPGe detectors (the original 1 kg detector and the three new 2.42 kg detectors) are housed in their copper cryostat endcaps inside the detector chamber, They are surrounded by the innermost lead and borated-polyethylene shield layers and by the plastic scintillator muon veto. The copper cryostat arms connecting each cryostat endcap to its cryocooler outside the shield are visible along the four axes.}
    \label{fig:setup}
\end{figure}

To reach the low background levels required for a  \CEvNS~search, the \conusplus~high-purity germanium (HPGe) detectors are housed in a compact, multi-layer shield made of lead, polyethylene and borated polyethylene, enclosed in a stainless-steel frame, providing passive shielding against ambient gamma and neutron radiation. Two layers of plastic scintillator plates, each formed of 9 individual panels read out by 2--4 photomultiplier tubes (PMTs), are embedded in the shield and act as an active muon veto, tagging and rejecting muon-induced events. The inner detector chamber is continuously flushed with radon-free air to suppress the radon background. Each diode is mounted in an ultra-low-background copper cryostat and cooled by an  electrical cryocooler using a pulse tube and an automotive coolant for heat exchange. Combined, these measures reduce the total background by four orders of magnitude relative to an unshielded configuration.  More details are shown in \cite{CONUS:2024lnu}. Figure~\ref{fig:setup} shows the open shield with the four run 2 detectors installed in their copper cryostats, illustrating the endcap size difference between the new 2.42~kg diodes and the original 1~kg detector.

The active mass of each new detector is determined from the manufacturer's (Mirion) crystal dimensions and partial charge collection layer (PCCL) specifications. For one of the new large detectors, this is additionally validated by measuring the attenuation of gamma lines from a $^{241}$Am source placed at different positions around the crystal. As an example, for one of the new detectors the manufacturer-determined active mass is $2326^{+18}_{-33}$~g ($^{+0.77\%}_{-1.44\%}$), with the dominant contributions to the asymmetric uncertainty arising from the thickness of the lithium-diffused partial charge collection layer ($600^{+200}_{-100}$~$\mu$m) and the bottom passivated layer thickness (500--1000~$\mu$m). For the remaining original detector, C3, we instead rely on equivalent local attenuation measurements performed previously \cite{Bonet:2020ntx}. C3 was deliberately kept in place through the upgrade rather than replaced, in order to monitor for any change in background, for example introduced by some $^{60}$Co surface contamination~\cite{CONUS:2024vyx}. The spectra recorded before and after the upgrade were found to be consistent, confirming that no additional background was introduced during the intervention. One of the three new large detectors was found to be non-operational following installation, likely due to damage to its front-end ASIC, leaving a total active mass of $(5.60^{+0.03}_{-0.04})$~kg for run 2.

\section{\label{sec:datataking}Data acquisition}

Following the detector upgrade described in Sec.~\ref{sec:exp}, \conusplus~run 2 data acquisition started in February 2025. In this section, we describe the resulting exposure and the stability of the detectors and experimental conditions throughout the run.

\subsection{\label{sec:exposure}Exposure}

\begin{figure*}[htb]
    \centering
    \includegraphics[width=\textwidth]{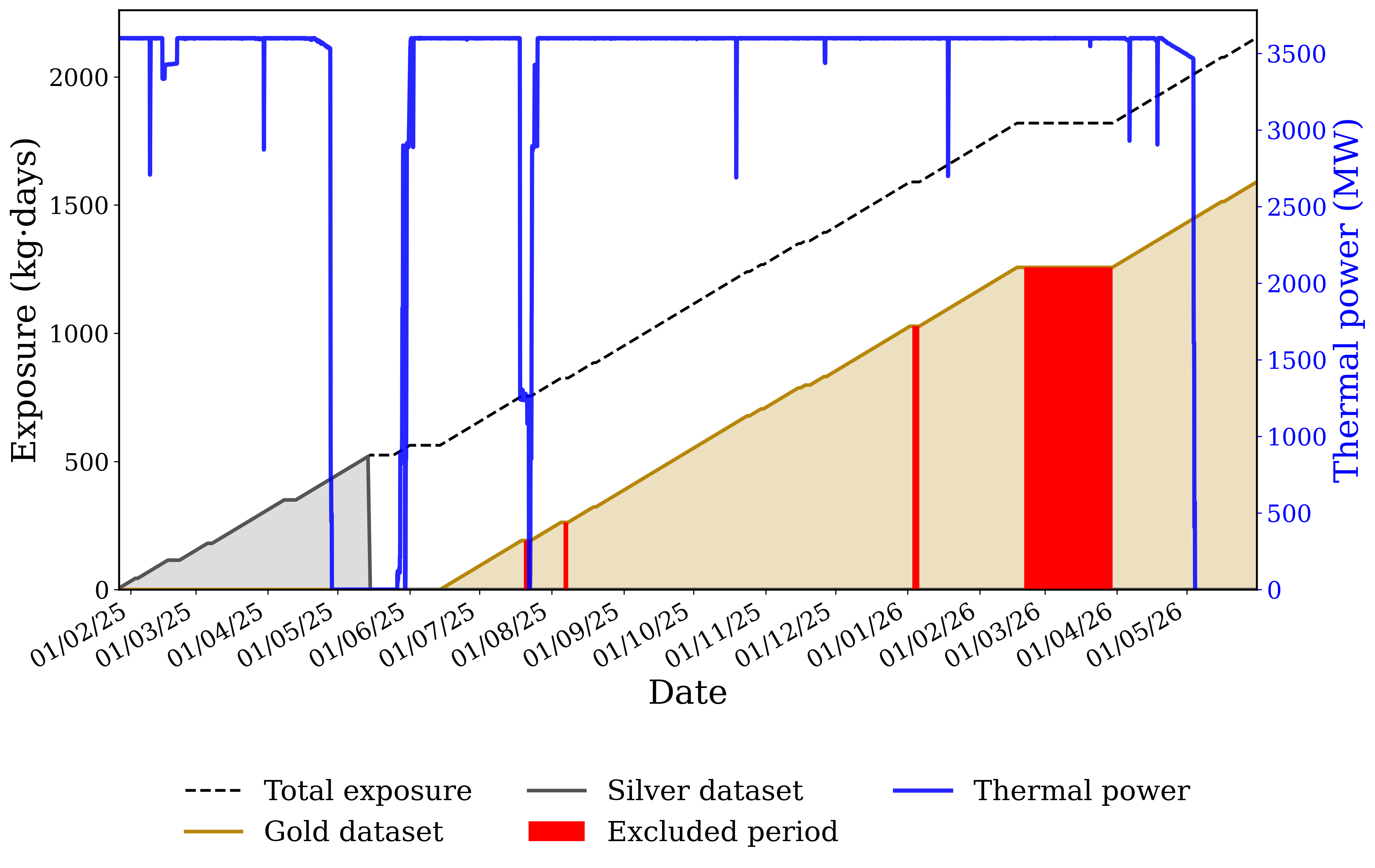}
    \caption{Accumulated exposure (left axis) and reactor thermal power (right axis, blue) as a function of time since the start of run 2. The black dashed line shows the total accumulated exposure. The silver dataset, collected before the May 2025 power cut, is shown in grey; the golden dataset, collected after the subsequent detector thermal cycle and used for the \CEvNS~analysis, is shown in gold. Periods excluded from the analysis due to instabilities in the radon flushing system or in the ground conditions are marked in red.}
    \label{fig:exposure}
\end{figure*}

Since the start of run 2, more than 2000~kg$\cdot$days of total exposure have been collected. Figure~\ref{fig:exposure} shows the accumulated exposure as a function of time, together with the reactor thermal power. Periods excluded from the analysis due to instabilities in the radon-suppression flushing system or in the electrical grounding conditions are marked in red.

Data acquisition was interrupted in May 2025 by a power cut, which caused a vacuum problem in the detector cryostats after warming up. The affected cryostats were pumped out again and the detectors underwent a full thermal cycle before resuming operation. The background rate and spectral shape recorded before and after this intervention were found to be compatible. However, after this intervention small differences in leakage current, pulser resolution, and noise rate were observed. We therefore define two data tiers. The \textit{golden} dataset comprises only the data collected after the recycle and is used for the \CEvNS~physics analysis presented in this paper. The \textit{silver} dataset contains the data collected before the power cut and is used, together with the golden dataset, for the energy calibration, as well as for detector specific studies.

After applying the event selection described in Sec.~\ref{sec:analysis} and accounting for detector dead time, the exposure used for the \CEvNS~analysis is 1296~kg$\cdot$days reactor-on and 86~kg$\cdot$days reactor-off.

\subsection{\label{sec:stability}Data stability} 

\begin{figure*}[htb]
    \centering
    \includegraphics[width=\textwidth]{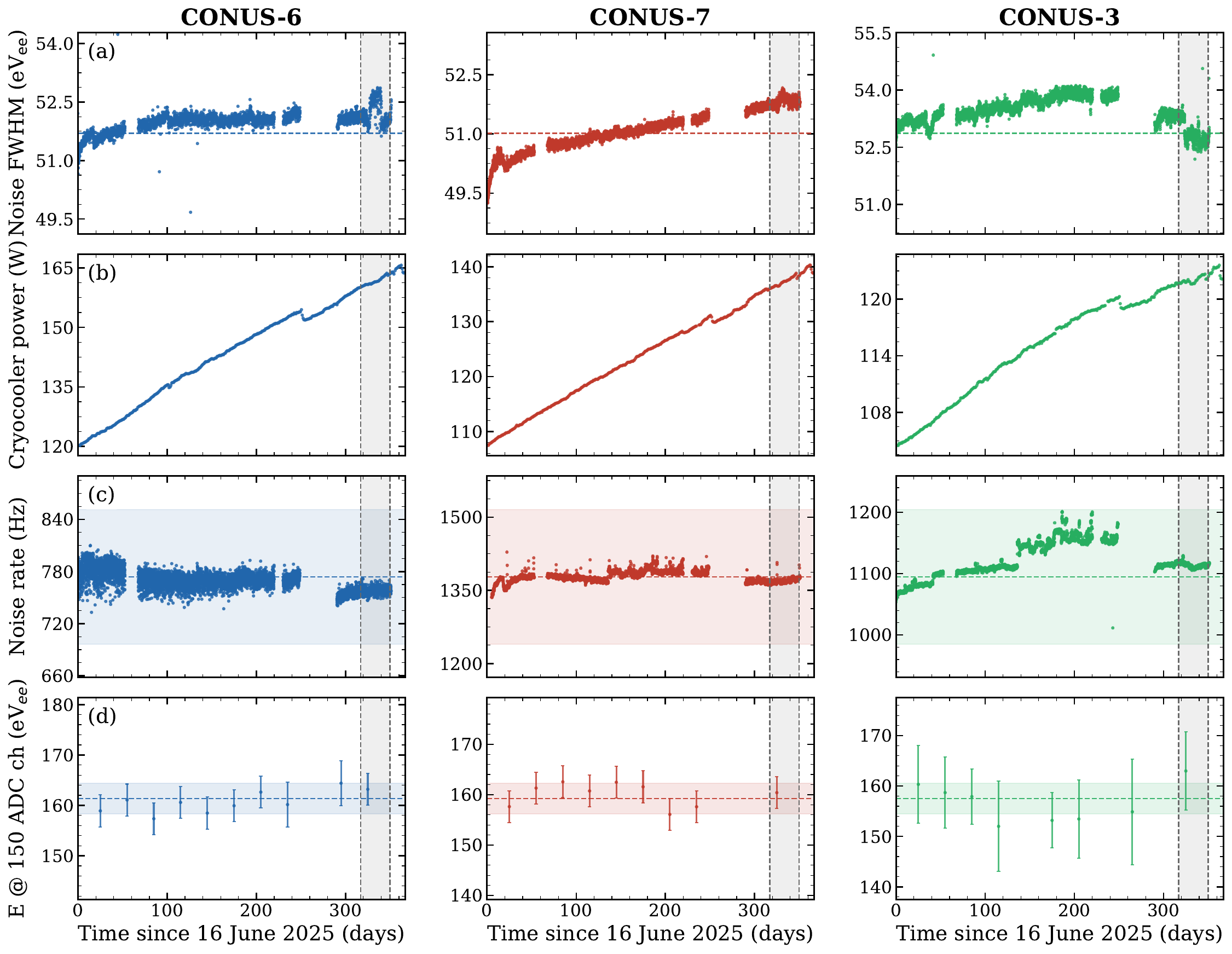}
    \caption{Stability of the detector operating conditions for C6 (left), C7 (center), and C3 (right): noise FWHM (a), cryocooler power (b), noise rate (c), and energy calibration monitored by tracking the calibrated energy of a fixed reference channel (150~ADC) over time (d). Dashed lines indicate the mean value over the shown period for each detector, and the colored bands the $\pm10\%$ variation of the noise rate (c) and the $\pm3$~eV$_{ee}$ energy-scale uncertainty (d). The shaded area marks the reactor-off period.}
    \label{fig:stability_det}
\end{figure*}

The stability of the detector operating conditions was monitored throughout run 2 for all detectors and is shown in Fig.~\ref{fig:stability_det}, with C6 (CONUS-6) on the left, C7 (CONUS-7) in the center, and C3 (CONUS-3) on the right. The room temperature is maintained stable with an air-conditioning system, kept at approximately 21$^{\circ}$C with variations below 0.5$^{\circ}$C throughout the run. The noise full width half maximum (FWHM) was stable within 2.0~eV$_{ee}$ for all detectors (a), with a slow increase over the run for C7. The cryocooler power (b) showed a gradual increase over the course of the run, attributed to a vacuum problem affecting the gaskets of the detectors. This increase is not correlated with the event rate in the \CEvNS~ROI, with a Pearson correlation coefficient below 0.05. The trigger rate on electronic noise (c) remained stable throughout the data-taking period, with variations within 10$\%$ (colored band), meeting the previously specified stability requirement~\cite{Ackermann:2024kxo}. The trigger efficiency, discussed further in Sec.~\ref{sec:det}, was likewise found to be stable throughout the run within the corresponding systematic uncertainty. The stability of the energy calibration was independently verified by monitoring the calibrated energy corresponding to ADC channel 150 (d), close to the analysis threshold. The energy was determined using the same $^{71}$Ge calibration lines described in Sec.~\ref{sec:det}, with the calibration computed in 30-day time windows instead of over the full exposure. The resulting values are consistent within their uncertainties with the energy-scale uncertainty of $\pm3$~eV$_{ee}$ (colored band), with no significant drift observed over the course of the run.

\begin{figure*}[htb]
    \centering
    \includegraphics[width=\textwidth]{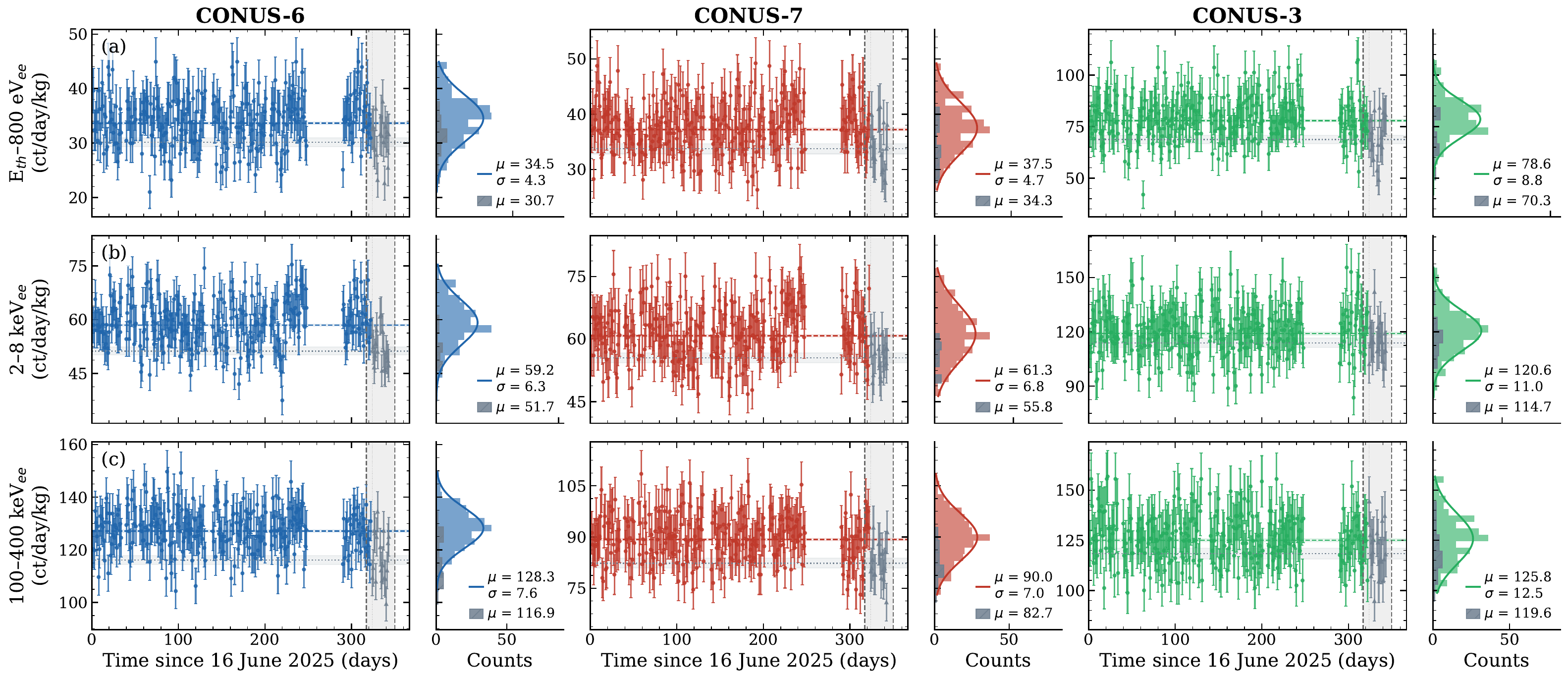}
    \caption{Stability of the background rate for C6 (left), C7 (center), and C3 (right) in the \CEvNS~ROI, evaluated after unblinding (a), the 2--8~keV$_{ee}$ range (b), and the 100--400~keV$_{ee}$ region (c), shown separately for reactor-on (colored) and reactor-off (gray) periods. Dashed and dotted lines indicate the mean values for reactor-on and reactor-off periods, respectively. The distributions of the daily rates are shown on the right of each panel, with a Gaussian fit for the reactor-on data. The shaded area marks the reactor-off period.}
    \label{fig:stability_bkg}
\end{figure*}

The background rate was monitored in three energy regions, separately for reactor-on and reactor-off periods, as shown in Fig.~\ref{fig:stability_bkg}: the \CEvNS~ROI below 800~eV$_{ee}$ (a), the 2--8~keV$_{ee}$ range (b), and the high-energy region between 100 and 400~keV$_{ee}$ (c), where the rate in the ROI was evaluated only after unblinding (see Sec.~\ref{sec:likelihood}). The rates were found to be stable throughout the data-taking period for all detectors, as also shown by the distributions of the daily rates on the right of each panel. The stability of the 100--400~keV$_{ee}$ region also demonstrates the efficiency of the radon flushing system, as no significant variations are observed. This was further verified by checking the stability of the three main radon-daughter $^{214}$Pb gamma-lines at 241.9, 295.2, and 351.9~keV. During reactor-off periods, the background rate is systematically lower by about 5--10\%, due to the additional overburden provided by the reactor drywell lid positioned above the experimental room during these periods. This effect is discussed in more detail in Sec.~\ref{sec:bkgmodel}.

\section{\label{sec:det}Detector performance}

The performance of the three run 2 detectors used in the \CEvNS~analysis (C6, C7, and C3) was characterized in terms of trigger efficiency, energy resolution, energy threshold, and energy-scale calibration and linearity, following the same general methodology as described for the \conusplus~setup in \cite{CONUS:2024lnu}.

\subsection{Energy calibration and linearity}
\label{sec:calibration}

The energy scale of each detector is established with a linear calibration, $E = a + b\cdot\mathrm{ch}$, where $\mathrm{ch}$ is the digitized pulse amplitude. Since low-energy $\gamma$-rays from an external source cannot penetrate efficiently the copper cryostat surrounding the HPGe diodes, the calibration instead exploits characteristic atomic de-excitation lines emitted internally within the crystals following electron capture decays of cosmogenically activated isotopes ($^{65}$Zn, $^{68}$Ga, $^{68}$Ge, $^{71}$Ge). These decays produce two dominant lines in the low-energy spectrum: a K-shell line at 10.367~keV and an L-shell line at 1.298~keV, both emitted by $^{68}$Ge and $^{71}$Ge. Additional, weaker K-shell lines are produced by $^{68}$Ga at 9.659~keV and by $^{65}$Zn at 8.979~keV. The L-shell lines of $^{68}$Ga (1.19~keV) and $^{65}$Zn (1.10~keV) lie close to the dominant L-shell line of $^{68}$Ge and $^{71}$Ge and partially overlap with it at the resolution of the \conusplus~detectors. Since their intensities are much lower than that of the 1.30~keV line, the observed peak structure can be well described by a single Gaussian. The resulting energy resolution at the K-shell and L-shell atomic de-excitation lines for each detector is summarized in Table~\ref{tab:calib}.

\begin{table}[htb]
    \centering
    \caption{FWHM at 1.3~keV and 10.4~keV for each detector, obtained from the energy calibration.}
    \label{tab:calib}
    \setlength{\tabcolsep}{3pt}
    \renewcommand{\arraystretch}{1.7}
    \begin{tabular}{lcc}
        \hline\hline
        Detector & FWHM @ 1.3~keV (eV) & FWHM @ 10.4~keV (eV) \\
        \hline
        C6 & $83.8 \pm 2.16$ & $168 \pm 0.89$ \\
        C7 & $82.0 \pm 1.79$ & $168 \pm 0.82$ \\
        C3 & $80.4 \pm 4.16$ & $163 \pm 1.51$ \\
        \hline\hline
    \end{tabular}
\end{table}

Among these isotopes, the $^{71}$Ge lines serve as the main calibration reference, as $^{71}$Ge is continuously produced in situ by muon-induced neutrons interacting with the \conusplus~shield~\cite{CONUS:2024lnu}, giving a steady K-shell rate of approximately 40~counts~kg$^{-1}$~d$^{-1}$, roughly a factor of 7 higher than the L-shell rate~\cite{CONUS:2024lnu}. In contrast, the $^{65}$Zn, $^{68}$Ga, and $^{68}$Ge activity is produced on surface and decays away over the course of the run, with $^{68}$Ge and $^{65}$Zn half-lives of 270.95 and 244.01~days. Considering the full run 2 exposure, combining both the silver and gold datasets, propagating the calibration uncertainties to the ROI yields an energy-scale uncertainty at threshold of approximately 1.6~eV$_{ee}$ for C6 and C7, and 2.5~eV$_{ee}$ for C3, in agreement with the expectations from the nominal $^{71}$Ge production rate~\cite{SanchezGarcia:2026nvr}. The smaller uncertainty for C6 and C7 is due to the higher statistics of their calibration lines, from both the larger active mass, which increases the number of in-situ $^{71}$Ge decays, and their more recent cosmogenic activation, which increases the $^{68}$Ge and $^{65}$Zn activity. The stability of the calibration coefficients was monitored throughout run 2 by repeating the fit independently for consecutive 30-day periods, as shown in Fig.~\ref{fig:stability_det}. The resulting scatter of the calibrated energy across these periods is at the level of 2-3~eV$_{ee}$ for the three detectors. For the systematic uncertainty on the energy scale, a common, conservative value of $\pm3$~eV$_{ee}$ is adopted for all three detectors in the likelihood analysis (Sec.~\ref{sec:likelihood}).

\begin{figure}[htb]
    \centering
    \includegraphics[width=\columnwidth]{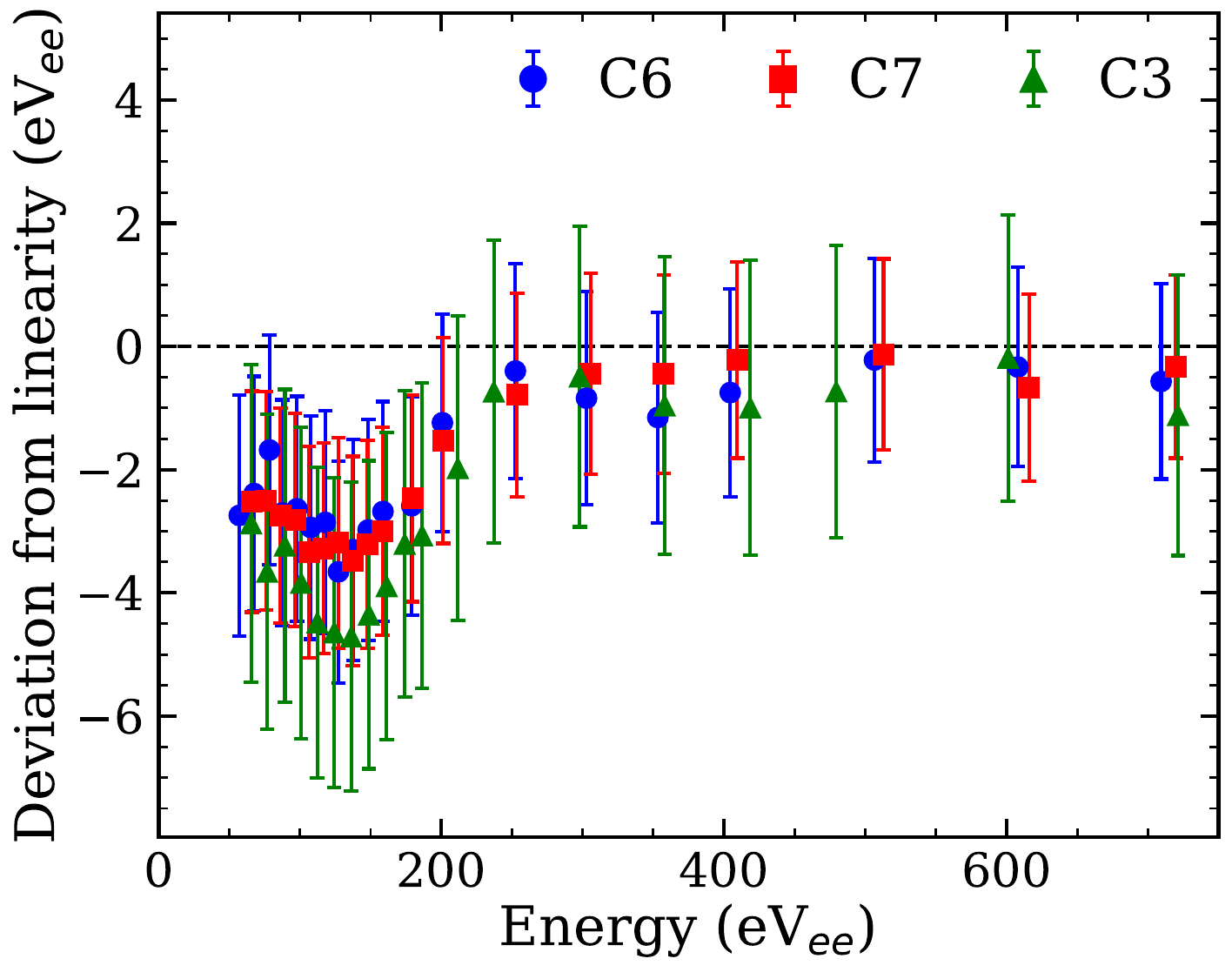}
    \caption{Deviation from linearity as a function of energy for detectors C6, C7, and C3, determined from pulser injection measurements.}
    \label{fig:linearity}
\end{figure}

Since no clear calibration lines are available in the spectrum below 1.3~keV$_{ee}$, the linearity of the energy reconstruction in the ROI is instead verified by injecting artificial pulser signals across the ROI and measuring the deviation between the reconstructed and injected energy. This pulser-based approach was validated in~\cite{SanchezGarcia:2026nvr}, where a dedicated neutron activation campaign enhanced the M-shell line (158~eV) statistics sufficiently to resolve it directly, confirming that the pulser-corrected energy scale reproduces the expected M-shell position. In run 1, non-linear deviations from the DAQ chain were observed below 250~eV$_{ee}$~\cite{Ackermann:2025obx}, reaching a variation up to 15~eV$_{ee}$ around 160~eV$_{ee}$. For run 2, this non-linearity is substantially reduced through a combination of a longer trapezoidal rise time in the DAQ energy reconstruction and, as described in Sec.~\ref{sec:analysis}, an offline energy correction based on the waveform maximum position, bringing the maximum non-linearity down to approximately 4~eV$_{ee}$ for all detectors as shown in Figure~\ref{fig:linearity}.

\subsection{Detector response}
\label{sec:response}

\begin{figure}[htb]
    \centering
    \includegraphics[width=\columnwidth]{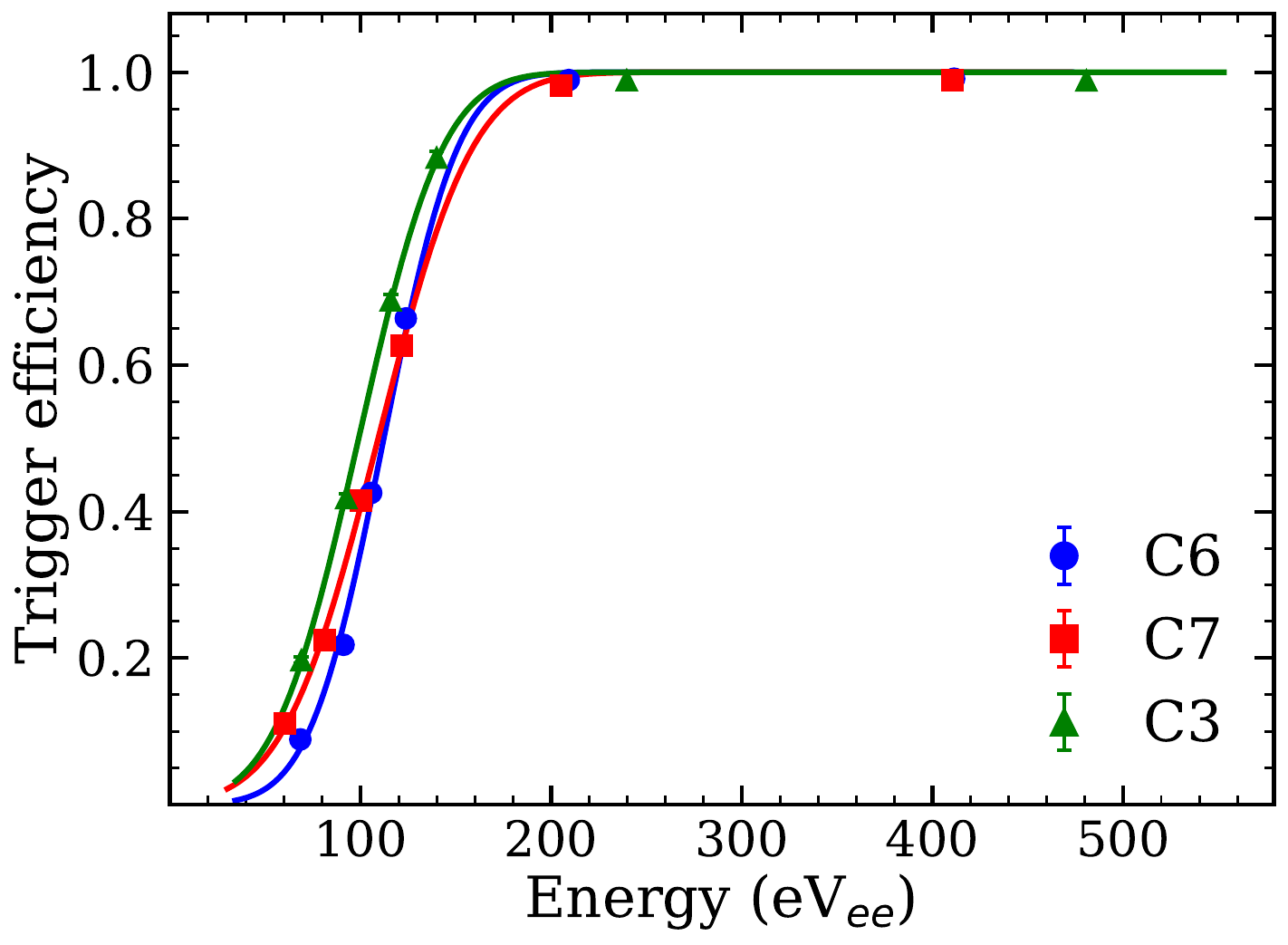}
    \caption{Trigger efficiency as a function of energy for detectors C6, C7, and C3, together with the corresponding best-fit curves using Eq.~\eqref{eq:trigeff}.}
    \label{fig:trigeff}
\end{figure}

The trigger efficiency as a function of energy is determined for each detector by injecting artificial pulser signals into the front-end electronics, with the same rise time as physical signals and step-wise varying amplitudes, allowing the detector response near the noise region to be finely scanned~\cite{Bonet:2023kob, CONUS:2024lnu}. The resulting curves are parametrized with a sigmoid function,
\begin{equation}
    \varepsilon_{trig} = 0.5 \cdot \left( 1 + \operatorname{erf} \left( \frac{E_{ee} - t_1}{t_2} \right) \right),
    \label{eq:trigeff}
\end{equation}
described by the parameters $t_1$ and $t_2$, where $t_1$ is the energy at which the trigger efficiency reaches 50\% and $t_2$ characterizes the width of the transition. Figure~\ref{fig:trigeff} shows the resulting trigger efficiency curves for C6, C7, and C3; all three detectors reach full efficiency down to  180~eV$_{ee}$. Table~\ref{tab:trigeff} summarizes the fitted $t_1$ and $t_2$ parameters for each detector, obtained from calibration measurements performed at regular intervals throughout run 2, together with their mean values and the root-mean-square (rms) spread across the run. The fitted parameters are found to be stable over time, and the corresponding rms values are used to build the covariance matrix of the trigger efficiency, which is propagated into the signal prediction and included in the likelihood (Sec.~\ref{sec:prediction}).

\begin{table}[htb]
    \centering
    \caption{Trigger efficiency parameters for each detector, with the energy at 50\% efficiency, $t_1$, the transition width, $t_2$,  with their rms over run 2.}
    \label{tab:trigeff}
    \setlength{\tabcolsep}{6pt}
    \renewcommand{\arraystretch}{1.7}
    \begin{tabular}{lcccc}
        \hline\hline
        Detector & $t_{1}$ (eV) & $t_{1}$ rms  (eV) &
        $t_{2}$ (eV) & $t_{2}$ rms (eV) \\
        \hline
        C6 & 110.4 & 0.84 & 47.8 & 2.35 \\
        C7 & 107.8 & 0.59 & 54.5 & 1.39 \\
        C3 & 99.1  & 0.82 & 46.3 & 1.36 \\
        \hline\hline
    \end{tabular}
\end{table}

The energy resolution of each detector is modeled as the standard Fano-noise resolution function, extended with an additional quadratic term,
\begin{equation}
    \sigma^2(E) = \sigma_{\mathrm{noise}}^2 + F\,\varepsilon\,E +
    \beta^2 E^2,
\end{equation}
where $\sigma_{\mathrm{noise}}$ is the electronic noise contribution, $F$ is the Fano factor, $\varepsilon$ is the average ionization energy, and $\beta$ is a dimensionless parameter accounting for pulse-shape and rise-time variability caused by inhomogeneities in the drift field. As noted in~\cite{SanchezGarcia:2026nvr}, this quadratic term has been used by other experiments~\cite{SuperCDMS:2018gro} to describe such effects. A dedicated precision calibration campaign performed with one of the new \conusplus~detectors, using neutron-activated $^{71}$Ge atomic de-excitation lines, demonstrated that this extended model provides a better description of the sub-keV energy resolution than the standard linear model, bringing the fitted Fano factor into better agreement with literature values~\cite{SanchezGarcia:2026nvr}. The electronic noise contribution, $\sigma_{\mathrm{noise}}$, is measured by injecting pulser signals directly into the front-end electronics and fitting the resulting pulser peak, giving FWHM values of ($52\pm1$)~eV$_{ee}$, ($51\pm1$)~eV$_{ee}$, and ($53\pm1$)~eV$_{ee}$ for C6, C7, and C3, respectively, consistent with the noise stability discussed in Sec.~\ref{sec:stability}. The $\sigma_{\mathrm{noise}}$ is constrained to the pulser-measured value. The resulting fit parameters for each detector are summarized in Table~\ref{tab:resolution}.

\begin{table}[htb]
    \centering
    \caption{Resolution model parameters for each detector: electronic noise contribution $\sigma_{\mathrm{noise}}$, Fano factor $F$, and $\beta$ parameter. The fitted parameters are consistent across all three detectors within their respective uncertainties.}
    \label{tab:resolution}
    \setlength{\tabcolsep}{6pt}
    \renewcommand{\arraystretch}{1.7}
    \begin{tabular}{lccc}
        \hline\hline
        Detector & $\sigma_{\mathrm{noise}}$ (eV) & $F$ &
        $\beta$ ($\times10^{-3}$) \\
        \hline
        C6 & $52.1 \pm 1.0$ & $0.134 \pm 0.010$ & $1.6 \pm 1.0$ \\
        C7 & $51.7 \pm 1.1$ & $0.130 \pm 0.010$ & $1.9 \pm 0.9$ \\
        C3 & $53.5 \pm 1.1$ & $0.126 \pm 0.011$ & $1.6 \pm 1.1$ \\
        \hline\hline
    \end{tabular}
\end{table}

Two independent criteria set the energy threshold of the \CEvNS~analysis window for each detector independently. First, a minimum 20$\%$ trigger efficiency must be reached, which happens down to 70~eV$_{ee}$. Second, the extrapolated contribution of the electronic noise peak must contribute less than 10$\%$ of the predicted \CEvNS~rate at the edge of the signal region~\cite{CONUS:2024lnu}. 

\begin{figure}[htb]
    \centering
    \includegraphics[width=\columnwidth]{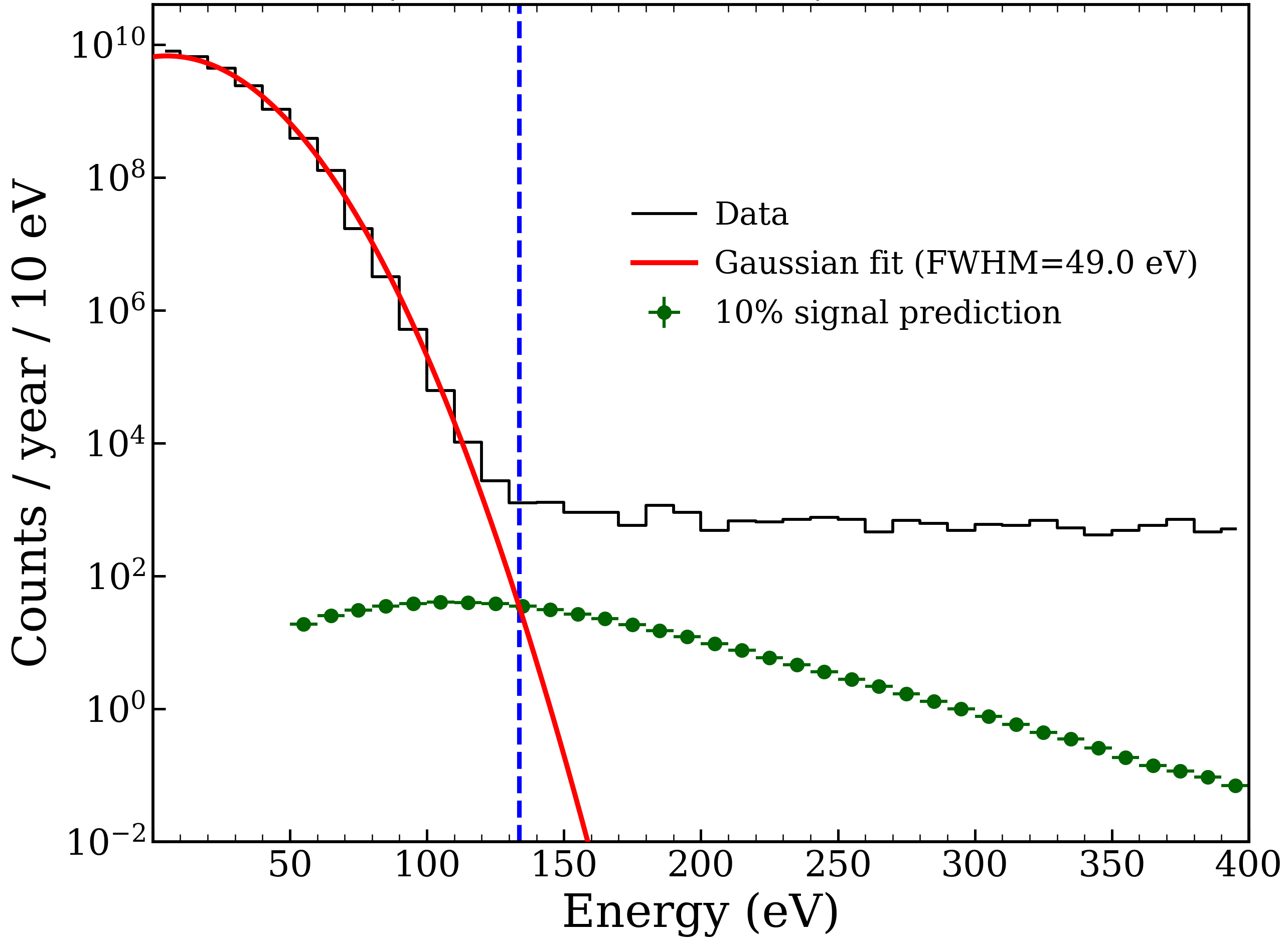}
    \caption{Example energy threshold determination, shown here for detector C7. The measured spectrum (black) is compared to a Gaussian fit of the electronic noise peak (red) and the 10$\%$ \CEvNS~signal prediction (green points); the threshold (dashed blue line) is set where the noise contribution falls above 10$\%$ of the signal prediction.}
    \label{fig:threshold}
\end{figure}

As an example, Fig.~\ref{fig:threshold} illustrates this determination for one detector, showing the measured noise peak, its Gaussian fit, and the 10~$\%$ of signal prediction used to identify the crossing point. The signal prediction assumes Lindhard quenching with $k = 0.162 \pm 0.004$~\cite{Bonhomme:2022lcz}. Applying this criterion to each detector yields threshold values of 160~eV$_{ee}$, 140~eV$_{ee}$, and 170~eV$_{ee}$ for C6, C7, and C3, respectively. The threshold of C3 is slightly higher than the 160~eV$_{ee}$ adopted in run 1~\cite{Ackermann:2025obx}, due to a slightly worse noise resolution, which we attribute to the increased leakage current after the thermal cycle. Together with the energy calibration established above, these detector-response parameters are used in the sensitivity studies and in
the \CEvNS~signal extraction presented in Secs.~\ref{sec:sensitivity} and~\ref{sec:result}, respectively.

\section{\label{sec:analysis}Data analysis}

We apply a series of selection cuts to reject non-physical and background-like events prior to the \CEvNS~likelihood analysis. These include a veto following each preamplifier reset, a time-difference cut targeting microphonics-induced events, a set of general waveform-quality cuts, and a muon-veto anticoincidence cut. In addition, a pulse-shape discrimination (PSD) study based on the pulse risetime and the current-amplitude-over-energy (A/E) parameter is performed. It is not applied as a selection cut, but used to validate the surface and multi-site event contributions predicted by the background model.

\subsection{\label{sec:cuts}Selection cuts}

Cosmic-ray muons are tagged with the two plastic scintillator veto systems described in Sec.~\ref{sec:exp}. A coincidence window of 450~$\mu$s between the veto and the HPGe signal is applied. This anticoincidence cut achieves a background rejection efficiency above 99\%. The total trigger rate of the muon veto is approximately 270~Hz, of which about 170~Hz originates from muons and the remaining 100~Hz from ambient gamma-rays induced signals; the trigger threshold of the veto is chosen to optimize the background rejection achieved by the anti-coincidence cut rather than to isolate muon-only events. This results in a $\sim$12\% dead time. More details about the muon veto performance are provided in~\cite{CONUS:2024lnu}.

The transistor-reset preamplifier (TRP) of each HPGe detector resets whenever the output ramp reaches the limit of its dynamic range. This ramp is driven mainly by the detector's leakage current, but is also incremented by the charge collected in each detected event; the resulting reset rate is 5 to 20~Hz depending on the detector. Immediately after a reset, the baseline is unstable, so a veto window of 1.5 to 2~$\mu$s depending on the detector, is defined following each reset during which no event is accepted. This veto suppresses events at low energy, becoming negligible above 5~keV$_{ee}$, and corresponds to an additional dead time of 1\% to 3\%. Since the TRP-reset veto window and the muon-veto coincidence window can overlap in time, the combined dead time from both cuts is computed using a rolling-window approach that merges overlapping vetoed intervals, avoiding double counting. This combined dead time is between 12-14\% depending on the detector, and constitutes the dominant contribution to the total exposure loss from the selection cuts. Events are further rejected based on the time difference to the preceding event in the same channel, following the time-difference distribution (TDD) method described in~\cite{Bonet:2020ntx}, in order to remove microphonic-induced events.

\begin{figure}[htb]
\centering
\includegraphics[width=\columnwidth]{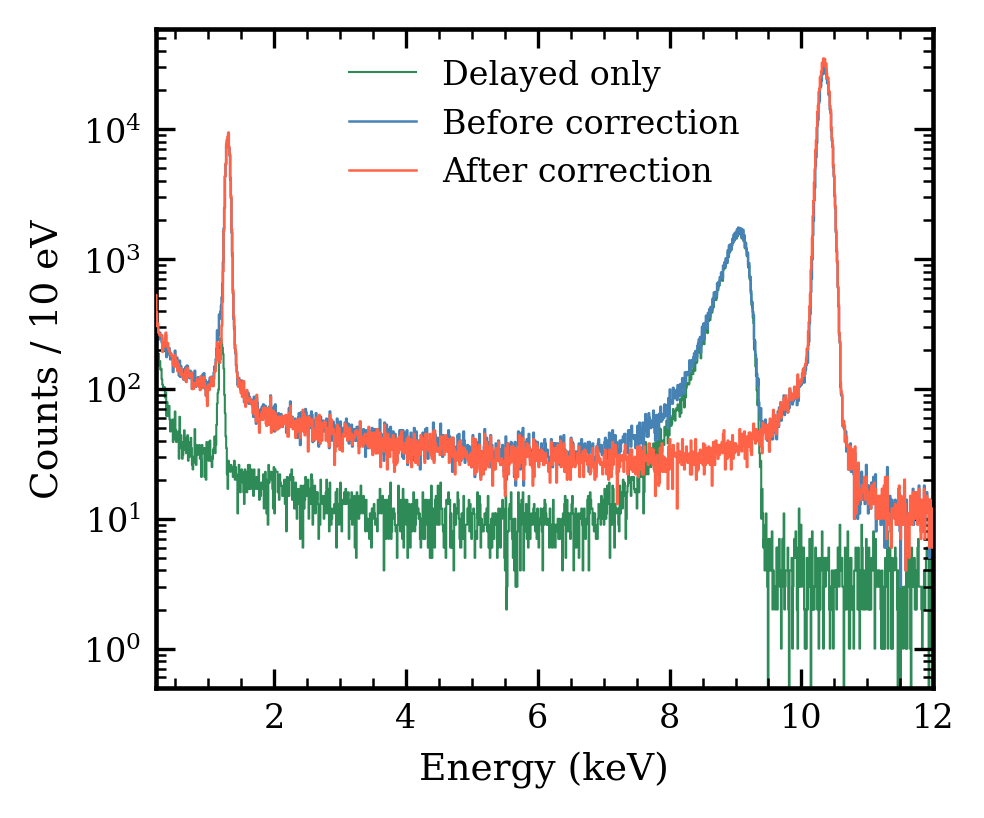}
\caption{Energy spectrum before (blue) and after the trigger-position correction (red), illustrating its effect on the $^{71}$Ge K-shell atomic de-excitation peak at $\sim$10.4~keV. The energy of the identified delayed events (green) is corrected considering the delay with respect to the expected trigger position. The example shown corresponds to detector C8 at MPIK, chosen for illustration because its larger irradiation-induced event rate in the $^{71}$Ge peaks makes the effect of the correction, and its recovery of the peak, more clearly visible than in the statistics available for the run 2 detectors.}
\label{fig:mshell_correction}
\end{figure}

A fraction of about 10\% of the remaining pulses, observed in all three detectors, shows a rising edge delayed relative to the trigger position. These events fall into two distinct categories. The majority show a systematic delay of the order of 1~$\mu$s in the pulse rise relative to the trigger position, leading to an underestimation of the reconstructed energy of up to 7\%. This effect is believed to be related to the triangular discrimination used in the DAQ trigger and is under further investigation. It is corrected by estimating the delay and the corresponding missing energy for each pulse individually, based on the trapezoidal-filter risetime and flat-top length of each detector. The remaining delayed events arise from random coincidences between a noise event and a physics pulse arriving later, in which the DAQ triggers on the noise event and the energy of the subsequent physics pulse is not properly reconstructed. As in the previous case, this is corrected by estimating the delay and the corresponding missing energy, but events with large delays cannot be reliably estimated and are removed instead. The effect of the correction is illustrated in Fig.~\ref{fig:mshell_correction} using detector C8, operated at the underground laboratory of the Max-Planck-Institut f\"ur Kernphysik (MPIK) for \conusplus~R\&D, chosen because its larger irradiation-induced event rate in the $^{71}$Ge K-shell peaks makes the recovery of the peak more clearly visible than the statistics available for the run 2 detectors allow.

An additional set of waveform quality cuts is applied. The maximum of the waveform is required to be below the maximum dynamic range of the ADC, removing saturated events. The pre-trigger region, defined as the first $1~\mu$s of the waveform, is required to be flat and not decreasing relative to a preceding event, rejecting events piled up on the falling tail of a previous pulse. Events for which the maximum of the waveform occurs before the expected trigger position are also removed, as these correspond to waveforms that are in fact the tail of a previous event rather than a genuine trigger-aligned pulse. The full set of waveform-based cuts is validated through visual inspection of both the retained and the rejected waveforms. 

\subsection{\label{sec:veto}Muon veto}

Cosmic-ray muons crossing the setup or its surroundings can produce backgrounds in the germanium detectors, either directly or through the production of secondary neutrons, gamma-rays and electromagnetic cascades. These events are tagged using the two active plastic scintillator vetoes described in Sec.~\ref{sec:exp} and in~\cite{CONUS:2024lnu}, and rejected through the anticoincidence cut discussed in Sec.~\ref{sec:analysis}. Among all selection cuts applied in the analysis, the muon cut provides by far the largest background reduction, and it is therefore essential to verify that the veto performance remains stable over the full duration of the run.

\begin{figure*}[htb]
\centering
\includegraphics[width=\textwidth]{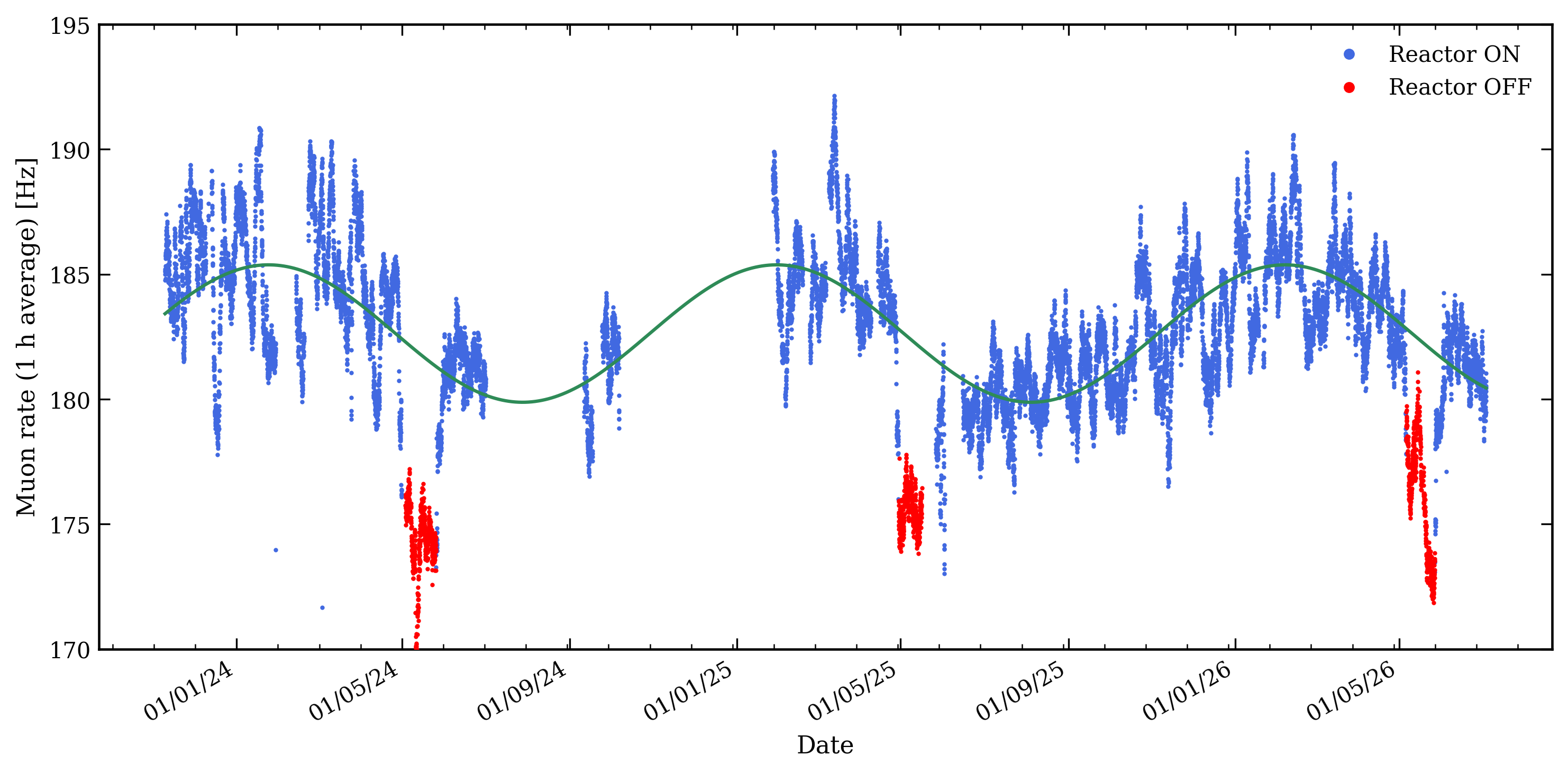}
\caption{Hourly muon rate measured by the \conusplus~muon veto over the full operation of \conusplus~at KKL, for reactor-on (blue) and reactor-off (red) periods; periods associated with detector upgrade operations or calibration campaigns are excluded. The reactor-on rate is systematically higher, reflecting the reduced overburden when the
reactor drywell head is not placed over the room. The solid line indicates the seasonal modulation expected from the temperature correlation discussed in the text.}
\label{fig:muon_rate_evolution}
\end{figure*}

As discussed in Sec.~\ref{sec:cuts}, the veto threshold is optimized for the overall background rejection of the anticoincidence cut and therefore also accepts ambient gamma radiation. To isolate the muon-induced component for the studies described below, a higher energy threshold is applied to achieve an almost pure muon rate of approximately 183~Hz for the full veto system, consistent with the expectation~\cite{CONUS:2024lnu}. Figure~\ref{fig:muon_rate_evolution} shows the evolution of the hourly muon rate over the full operation of \conusplus~at the Leibstadt nuclear power plant (Kernkraftwerk Leibstadt; KKL), covering both the run 1 and run 2, periods associated with detector upgrade operations or dedicated calibration campaigns are excluded from this dataset. During reactor-off periods, the muon rate is reduced by approximately 3\% relative to reactor-on periods, due to the positioning of the reactor drywell head on top of the room.

The muon rate is observed to vary during the year, correlated with atmospheric conditions. A fit of a sinusoidal function with a one-year period to the reactor-on rate (solid line in Fig.~\ref{fig:muon_rate_evolution}) yields a relative amplitude of $(1.51 \pm 0.02)\%$ with respect to the mean rate of 183~Hz, a period of 372~days, and a phase of 23 days, with the maximum reached at the end of January and the minimum at the end of July. This dependence is described using the effective-temperature formalism~\cite{Barrett:1952}. Since the atmosphere is non-isothermal, muon-flux variations at different heights are combined into a single effective temperature:

\begin{equation}
T_{\rm eff} = \frac{\int_0^\infty dX\, T(X) W(X)}{\int_0^\infty dX\, W(X)},
\end{equation}

where $X$ is the atmospheric depth and $W(X)$ is a weight function that depends on the threshold energy $E_{\rm thr}$, i.e.\ the minimum energy a muon needs to reach the detector, and on the zenith angle $\theta$ through the product $E_{\rm thr}\cos\theta$~\cite{Grashorn:2010}. For a uniform overburden, $E_{\rm thr}(\theta) \approx E_{\rm thr}(0)/\cos\theta$, so that $E_{\rm thr}\cos\theta$ is nearly independent of the zenith angle and can be evaluated from the vertical overburden alone, yielding $\langle E_{\rm thr}\cos\theta\rangle \approx 1.98$~GeV.

Hourly temperature profiles at 37 unevenly-spaced pressure levels (1--1000~hPa) are taken from the ERA5 reanalysis~\cite{ERA5} dataset of the European Centre for Medium-Range Weather Forecasts (ECMWF). The correlation study uses reactor-on data only, since the drywell head alters the overburden during reactor-off periods. The muon-rate variation is then fitted with:

\begin{equation}
\frac{I - \langle I \rangle}{\langle I \rangle} =
\alpha \frac{T_{\rm eff} - \langle T_{\rm eff} \rangle}{\langle T_{\rm eff} \rangle}
+ \beta (P - \langle P \rangle) + c,
\end{equation}

where $I$ is the hourly muon rate measured by the veto, $\langle I\rangle$ and $\langle T_{\rm eff}\rangle$ are the mean rate and mean effective temperature over the fitted period, $P$ and $\langle P\rangle$ are the ground pressure and its mean, and $c$ is a constant offset. The fit yields $\alpha = -0.7385 \pm 0.0005\,\text{(stat.)} \pm 0.0029\,\text{(syst.)}$ and $\beta = (-1.941 \pm 0.001\,\text{(stat.)} \pm 0.005\,\text{(syst.)}) \times 10^{-3}$~hPa$^{-1}$. The systematic uncertainties on $\alpha$ and $\beta$ are dominated by the accuracy of the ERA5 model and by the uncertainty in the weight-function parameters; the former is assessed by comparison with nearby meteorological stations, and the latter following the approach of~\cite{Alekseev:2021yya}.

\begin{figure}[htb]
\centering
\includegraphics[width=\columnwidth]{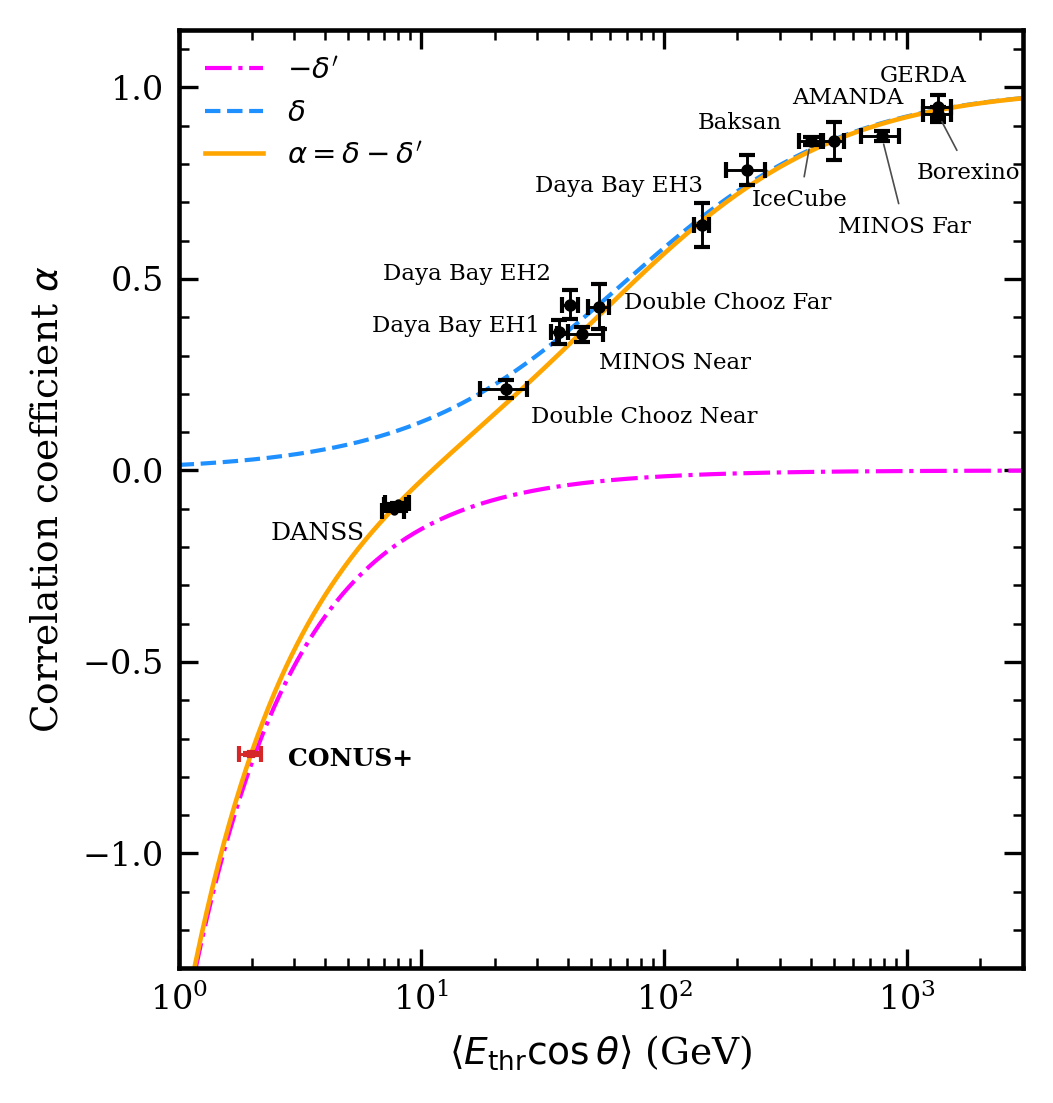}
\caption{Temperature correlation coefficient $\alpha$ as a function of $\langle E_{\rm thr} \cos\theta \rangle$,  compared to the standard temperature-correlation model including pion and kaon contributions, evaluated for a representative zenith angle (orange), and its two components: the positive contribution from enhanced meson decay at higher stratospheric temperature, $\delta$ (blue dashed), and the negative contribution from low-energy muon decay in the upper atmosphere, $-\delta'$ (magenta dash-dotted). For shallow experiments such as \conusplus, the negative contribution dominates while in deep underground experiments the positive contribution prevails. The \conusplus~measurement is shown together with other reactor and underground experiments: DANSS~\cite{Alekseev:2021yya}, Double Chooz~\cite{Abrahao:2017eyq}, Daya Bay~\cite{An:2018fyz}, MINOS Near~\cite{Adamson:2014vha}, MINOS Far~\cite{Adamson:2009zf}, Baksan~\cite{Andreyev:1991}, AMANDA~\cite{Bouchta:1999}, IceCube~\cite{Desiati:2011zz}, Borexino~\cite{Borexino:2018pev} and GERDA~\cite{GERDA:2016lhn}.}
\label{fig:alpha_comparison}
\end{figure}

Both coefficients are negative, but for different reasons. $\beta$ reflects the barometric effect, since at higher ground pressure the atmospheric column mass is larger, thus increasing muon absorption. $\alpha$ is negative because, at the shallow \conusplus~overburden, the effect of muons decaying higher in the atmosphere at higher stratospheric temperature dominates over the positive contribution from enhanced meson decay; the latter dominates in deep underground experiments. Figure~\ref{fig:alpha_comparison} compares the measured $\alpha$ with the prediction of the standard temperature-correlation model~\cite{Barrett:1952,Grashorn:2010} as a function of $\langle E_{\rm thr}\cos\theta\rangle$, together with results from other shallow-depth and deep underground experiments~\cite{Alekseev:2021yya,Abrahao:2017eyq,An:2018fyz,Adamson:2014vha,Adamson:2009zf,Andreyev:1991,Bouchta:1999,Desiati:2011zz, Borexino:2018pev, GERDA:2016lhn}. The model includes the contributions of muons from both pion and kaon decays, with critical energies $\epsilon_\pi = 114$~GeV and $\epsilon_K = 851$~GeV~\cite{Adamson:2009zf} and a kaon-to-pion ratio $r_K = 0.149$~\cite{Barrett:1952}, and is evaluated for a representative zenith angle. The \conusplus~value lies on the predicted curve, in the region where the negative term dominates. The observed muon-rate variations are thus fully explained by the known overburden and atmospheric effects, confirming stable veto operation.

\subsection{\label{sec:PSD}A/E and detector surface events}

The point-contact HPGe detectors exhibit a pulse shape that depends on the position of the energy deposition within the crystal. Events with a single energy deposition location within the detector (single-site events, SSE) produce a signal shape that is almost independent of where the interaction occurs, since the weighting potential is strongly localized near the read-out electrode; events with multiple energy deposition locations (multi-site events, MSE) instead produce a composite pulse shape. Separately, the charge-collection efficiency itself depends on the radial position of the energy deposition: it is zero within the lithium-diffused dead layer, rises smoothly from zero to unity across the transition layer as the drift field strengthens with increasing depth, and reaches full efficiency in the bulk volume, as described in more detail in Sec.~\ref{sec:deadlayer}. The energy of events interacting in the transition layer is therefore only partially and slowly collected, producing a systematically slower-rising pulse compared to bulk events. Following~\cite{Bonet:2023kob}, we refer to these as surface events, as opposed to bulk events, where charge collection is faster. For this reason, pulse-shape discrimination (PSD) helps to reduce the background, as \CEvNS~interactions are single-site, bulk events, while the background is more likely to be multi-site or to originate in the transition layer. This is exploited in two complementary techniques: the current-amplitude-over-energy (A/E) parameter targets the single-site/multi-site separation, and the pulse risetime targets the bulk/transition-layer separation, following the data-driven method developed for the CONUS detectors in~\cite{Bonet:2023kob}.

\begin{figure*}[htb]
\centering
\includegraphics[width=0.48\textwidth]{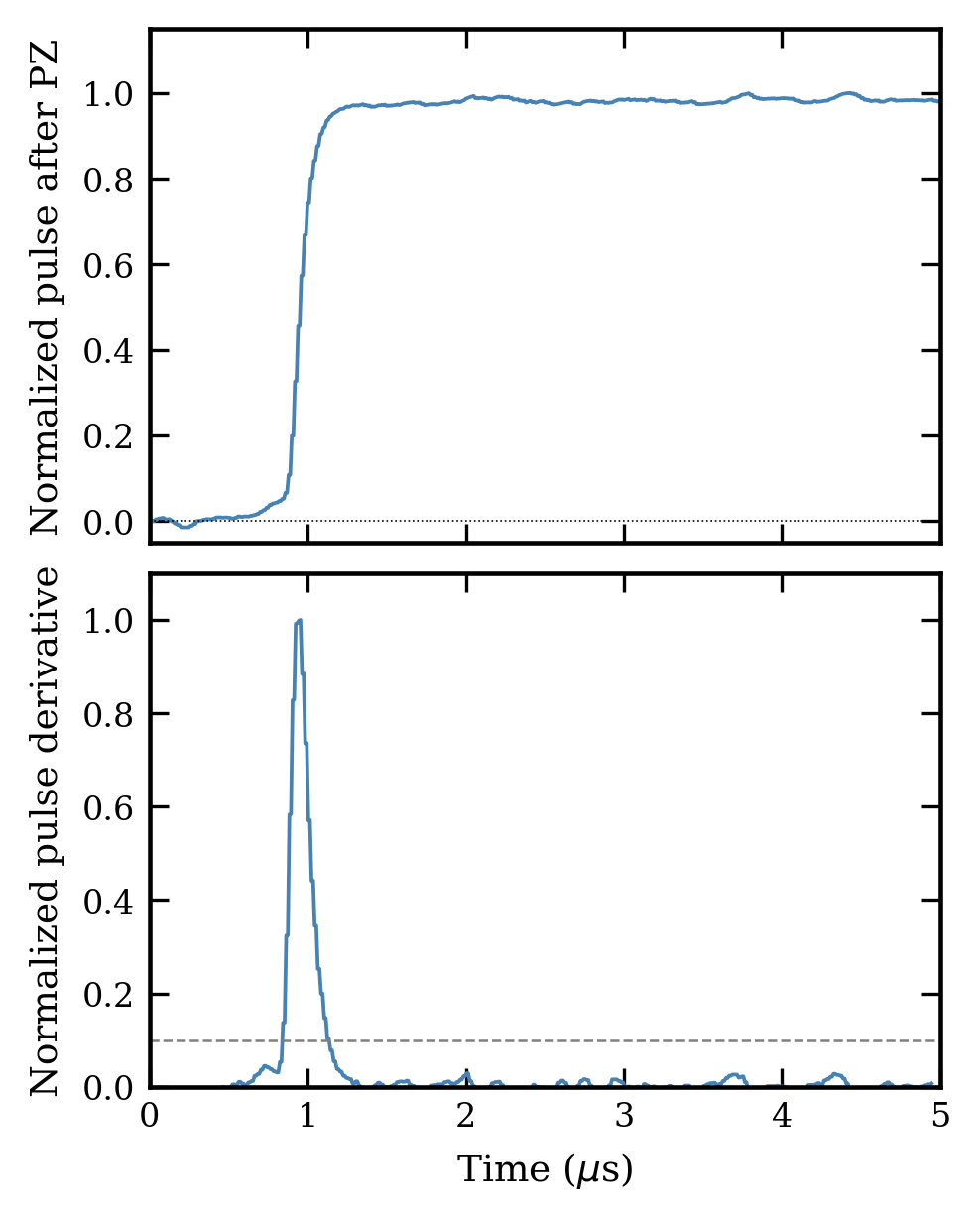}\hfill
\includegraphics[width=0.48\textwidth]{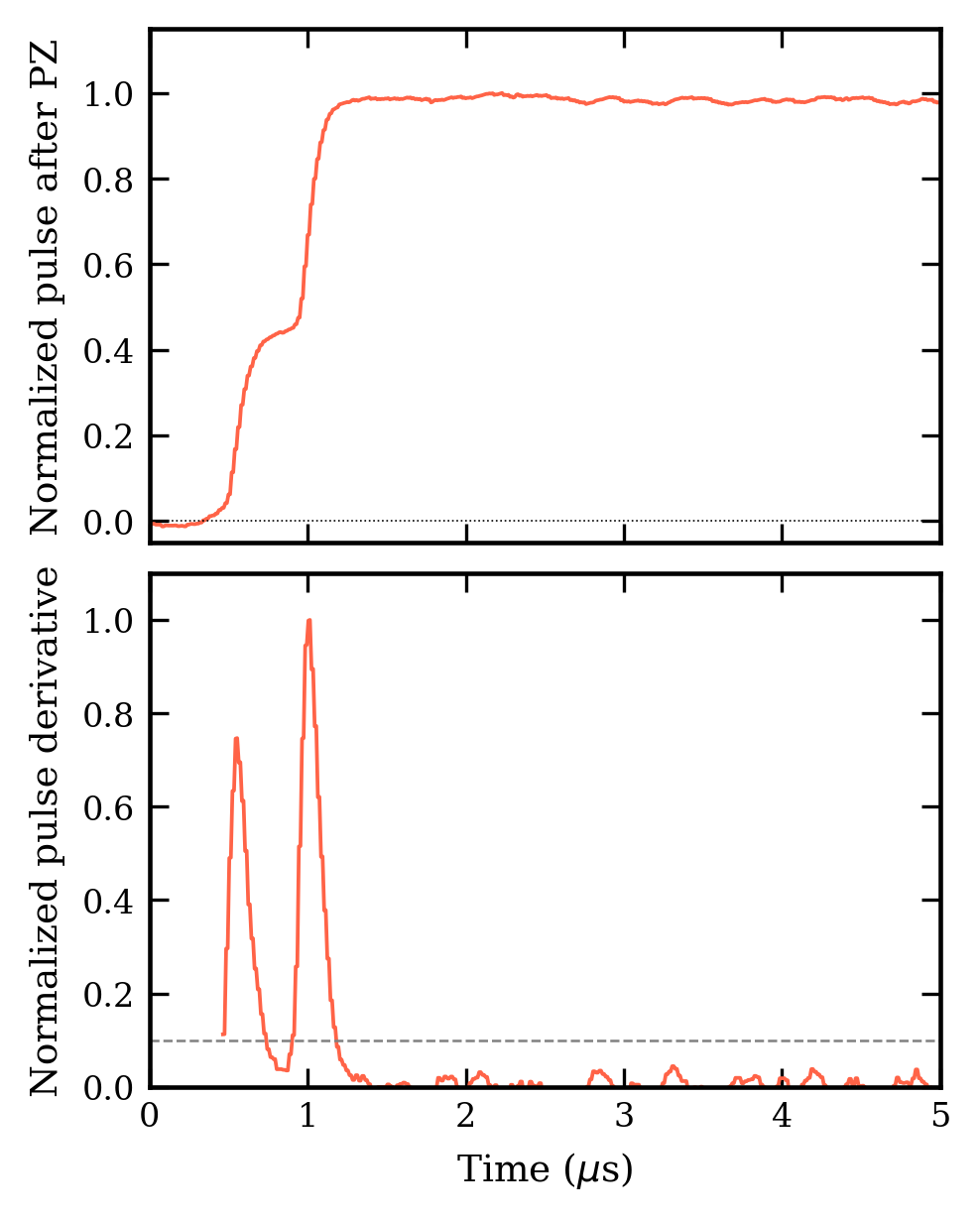}
\caption{Example normalized pulse waveforms after pole-zero correction (top panels) and their derivative (bottom panels), for a single-site event (left) and a multi-site event with two resolved current peaks (right), both with a reconstructed energy of about 5~keV$_{ee}$. The dotted line indicates the minimum peak height required.}
\label{fig:waveform_examples}
\end{figure*}

The A/E parameter is defined as the ratio of the maximum amplitude of the current pulse to the reconstructed energy of the event and was originally proposed for point-contact germanium detectors  in~\cite{Budjas:2009jm}. The current pulse is obtained as the time derivative of the waveform, and its maximum amplitude is divided by the reconstructed energy of the event, normalized such that single-site, bulk events cluster around $A/E \approx 1$; multi-site events instead exhibit a broader, systematically lower A/E, since their composite current pulse is distributed over more than one distinguishable charge-collection step rather than concentrated in a single peak. Multi-site events are identified from the number of resolved peaks in the current-pulse derivative. The waveform is first pole-zero (PZ) corrected, and its derivative is smoothed with an 80~ns moving average, peaks are then identified subject to minimum 10\% relative height and a 200~ns minimum separation. Figure~\ref{fig:waveform_examples} shows an example of a single-site event alongside a multi-site event, illustrating the characteristic single, sharp current peak of the former compared to the two distinguishable current peaks of the latter.

\begin{figure}[htb]
\centering
\includegraphics[width=\columnwidth]{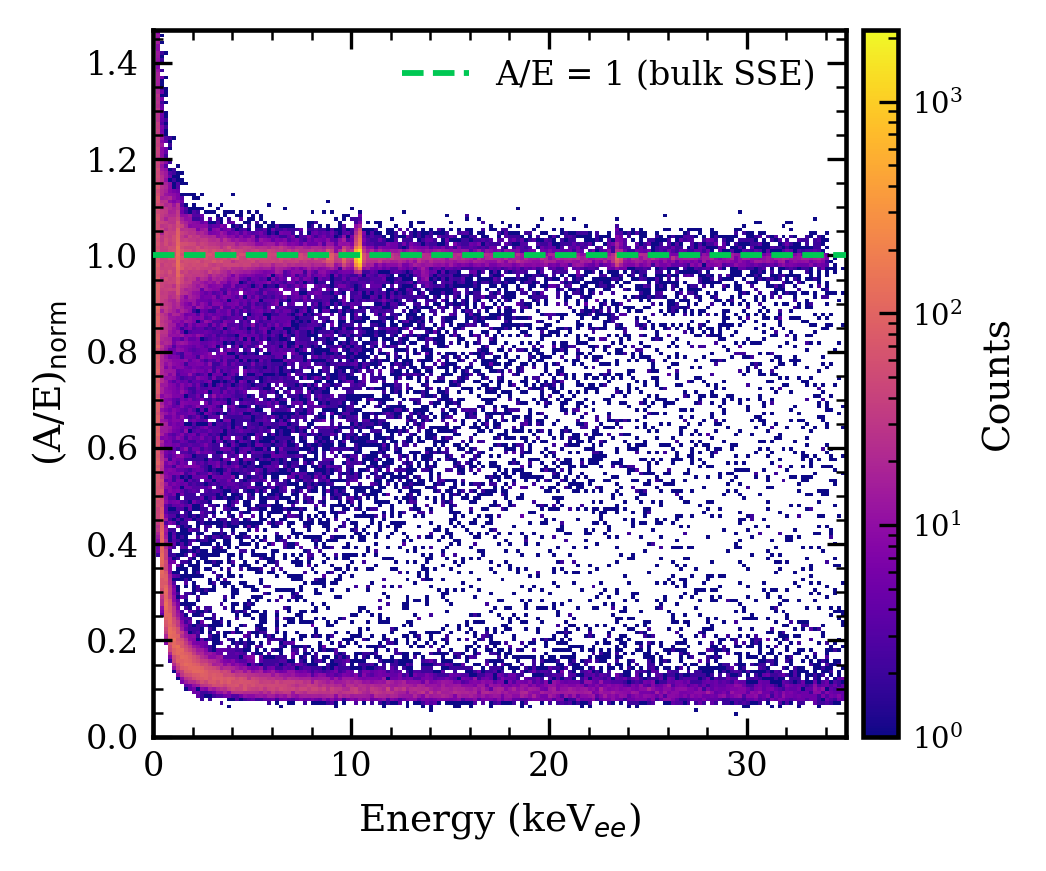}
\caption{Normalized A/E as a function of reconstructed energy for detector C3. Bulk single-site events cluster at A/E$\approx$1 (green dashed line), surface events populate the band at A/E$\approx$0.1, and multi-site events fill the region in between.}
\label{fig:AE}
\end{figure}

Figure~\ref{fig:AE} shows the resulting normalized A/E as a function of energy for C3. Three populations are visible: bulk single-site events forming a narrow band at A/E$\approx$1 including the atomic de-excitation lines of cosmogenic isotopes; surface events from the transition layer forming a band at A/E$\approx$0.1, whose slow charge collection reduces the current amplitude; and an intermediate continuum, populated mainly by multi-site events. The bulk and surface populations remain separated down to about 1--2~keV$_{ee}$, below which electronic noise broadens both bands and they merge.

\begin{figure*}[htb]
\centering
\includegraphics[width=0.48\textwidth]{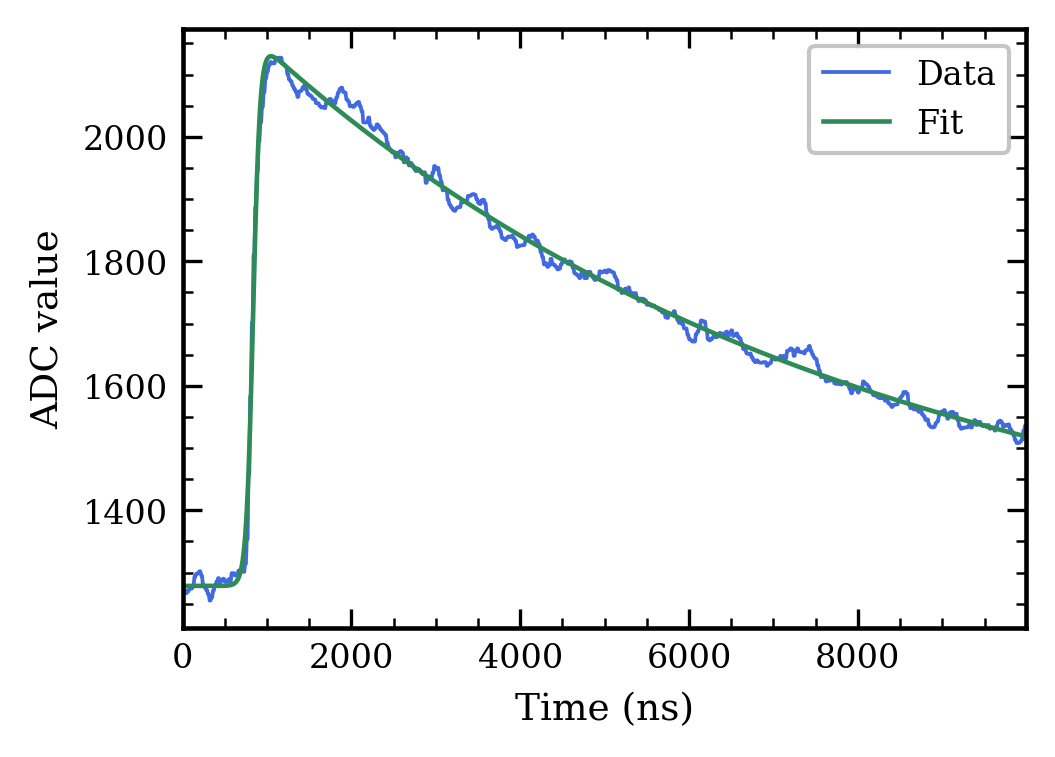}\hfill
\includegraphics[width=0.48\textwidth]{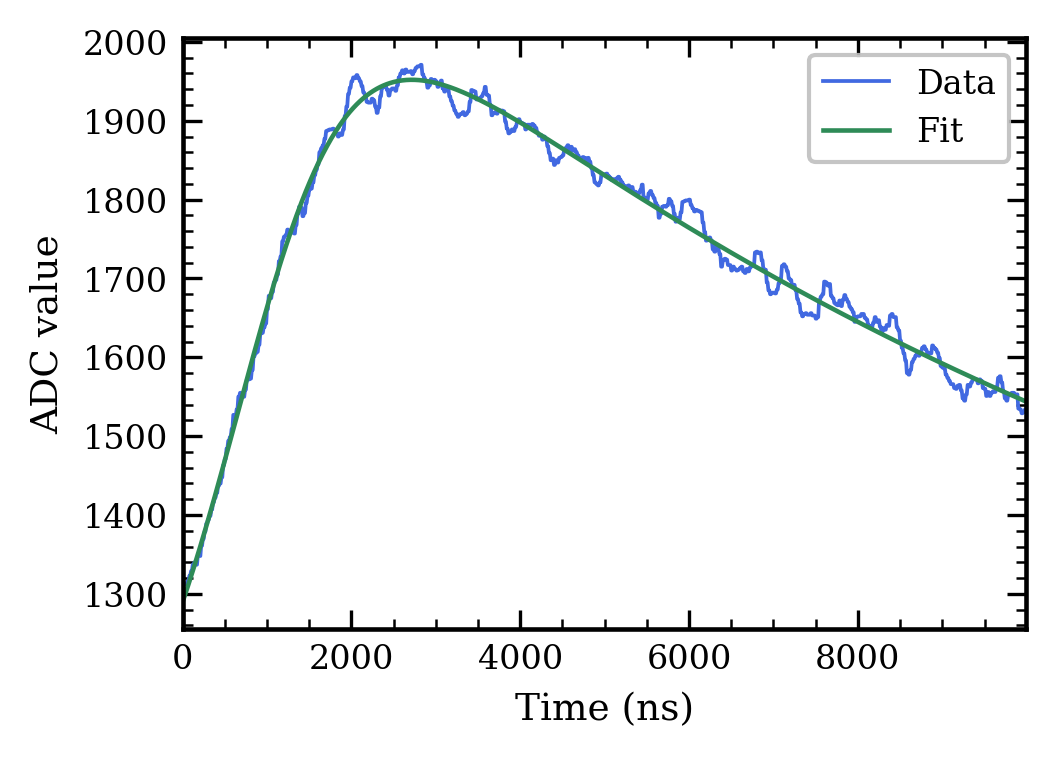}\\[4pt]
\includegraphics[width=0.48\textwidth]{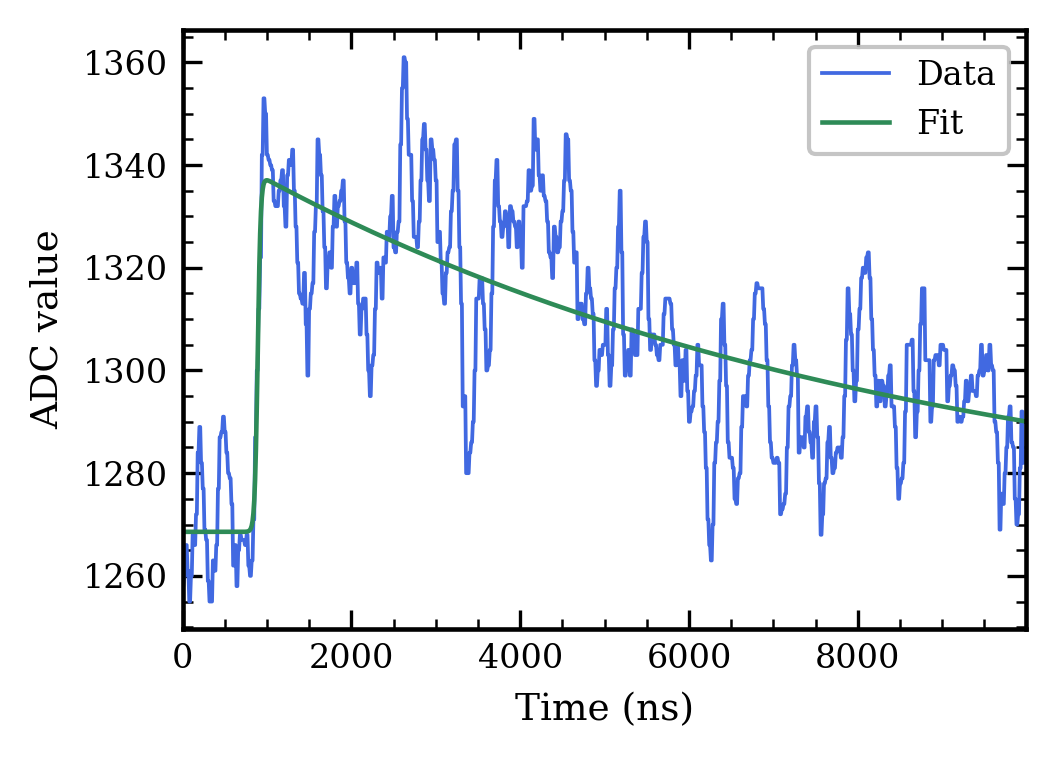}\hfill
\includegraphics[width=0.48\textwidth]{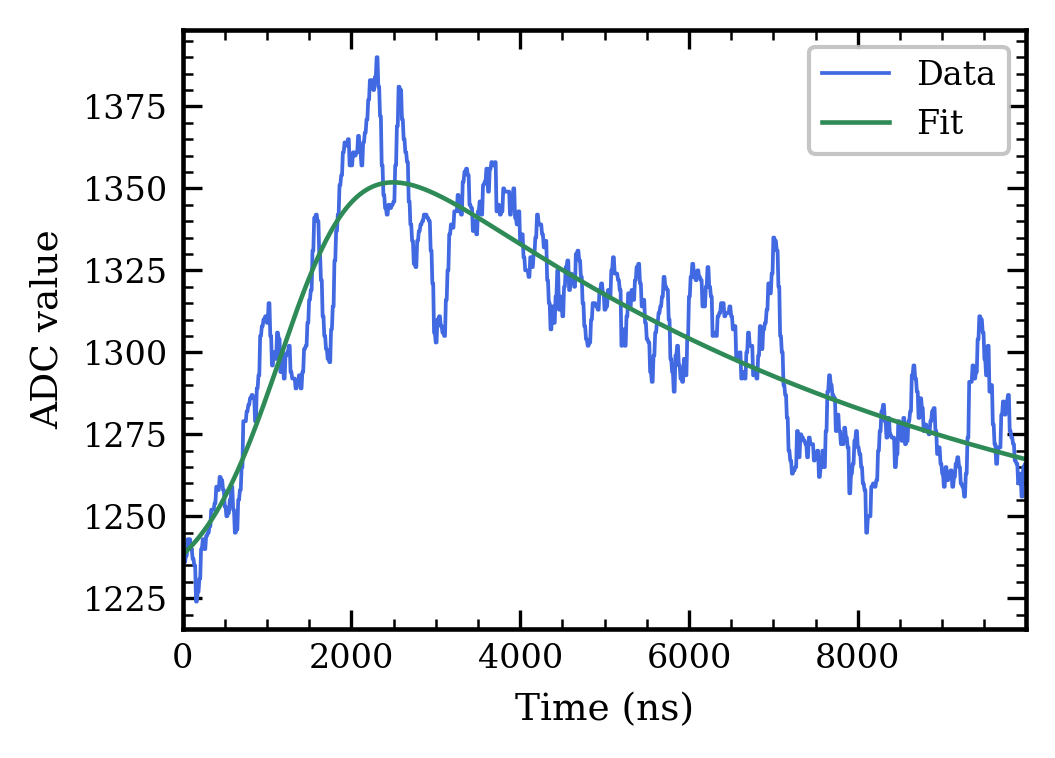}
\caption{Example waveforms and fits for bulk (left) and surface (right) events, at 2~keV$_{ee}$ (top) and 200~eV$_{ee}$ (bottom) energy ranges, illustrating the fast, single-step rise characteristic of bulk events compared to the delayed, slower rise of surface events.}
\label{fig:bulk_slow_examples}
\end{figure*}

Multi-site events, identified from the number of resolved peaks in the current pulse, are separated out before any further pulse-shape analysis, since the single-site risetime fit does not describe their multi-step waveform well, which would otherwise contaminate the bulk and transition-layer bands discussed below. No multi-site cut is applied to the data used in the \CEvNS~analysis. A simplified simulation of the average drift field in each detector is used to convert between the spatial separation of two energy depositions and the corresponding time separation of their current pulses, allowing the multi-site definition used in simulation to be compared with the time-separation-based classification applied to data. Between 7.5\% and 10\% of events are identified as multi-site in the 2 to 8~keV$_{ee}$ region. This multi-site population is compared to the expectation from a simplified simulation of gamma and neutron-induced background, which predicts a fraction of approximately 8\% was obtained for the 2.42~kg detectors in the same energy region. For the remaining single-site events, the pulse risetime is extracted by fitting each waveform with the function described in~\cite{Bonet:2023kob}:

\begin{equation}
f(t) = A_0 \left[ \tanh\!\left(\frac{t-t_0}{\tau}\right) + 1 \right]
\exp\left(-\tau_c (t-t_0)\right) + P_0,
\end{equation}

where $A_0$ is the pulse amplitude, $t_0$ the pulse onset time, $P_0$ the pre-trigger baseline, and $\tau$ the parameter of interest, related to the pulse risetime. The exponential term accounts for the AC coupling of the DAQ, with a characteristic decay time $\tau_c \sim (7.5-8)~\mu$s; being a constant of the system, $\tau_c$ is fixed in the fit to help convergence at low energy. For all detectors, bulk events yield a fitted risetime $\tau$ of 100~ns, corresponding to a $t_{10-90\%}$ risetime of 220~ns, while events in the transition layer can exhibit risetimes of several hundreds of nanoseconds.  Figure~\ref{fig:bulk_slow_examples} shows an example bulk and surface events, together with the corresponding risetime fit for each.

\begin{figure}[htb]
\centering
\includegraphics[width=\columnwidth]{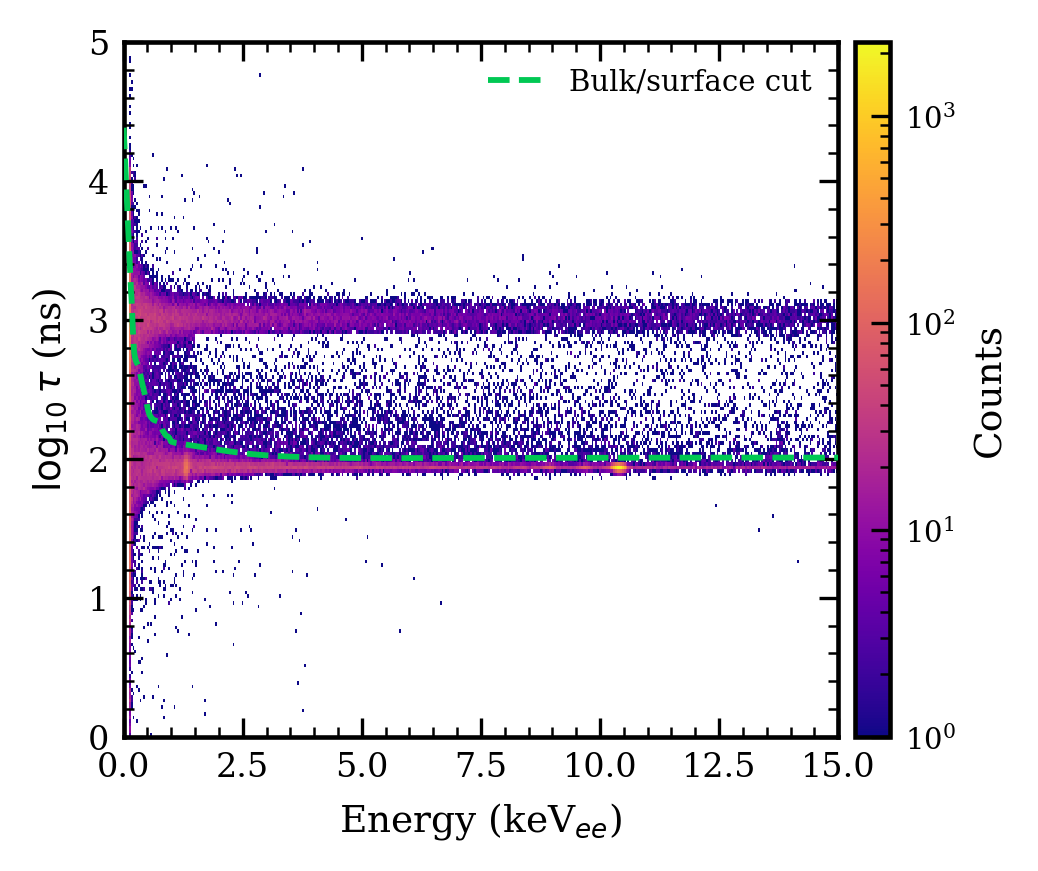}
\caption{Distribution of $\log_{10}(\tau)$ as a function of reconstructed energy for C3, showing the bulk and surface populations separated by the PSD threshold (green dashed line).}
\label{fig:risetime_vs_energy}
\end{figure}

Given the wide range spanned by $\tau$ between bulk and surface events, the distribution of $\log_{10}(\tau)$ as a function of energy is used throughout the following discussion, and is shown in Fig.~\ref{fig:risetime_vs_energy}. This distribution exhibits two populations, corresponding to bulk and surface events, with an intermediate region populated by multi-site and pile-up events, as well as events for which the risetime fit does not perform well, as described in~\cite{Bonet:2023kob}. For each detector and energy bin (20~eV$_{ee}$ wide below 1~keV$_{ee}$ and 200~eV$_{ee}$ above) the $\log_{10}(\tau)$ distribution is fitted with the sum of two Crystal Ball functions representing the bulk and surface populations, and a PSD threshold set at 3~$\sigma$ is used to estimate the surface event fraction as a function of energy. The resulting cut is shown with a green dashed line. Applying PSD as a selection cut in the analysis would require the full method proposed in~\cite{Bonet:2023kob}, including dedicated studies of the associated systematic uncertainties at low energies. This is not pursued here. Such a cut would reject about 20\% of the background in the ROI at a signal acceptance of about 95\%, resulting in only a small improvement in sensitivity. Instead, the surface fraction estimated from the risetime-based PSD threshold is compared to the expectation from a simplified Monte Carlo simulation of the neutron-induced background. In the 2--8~keV$_{ee}$ energy region, the observed surface fraction is approximately 35\% for all three detectors, exceeding the simulated fraction of approximately 15\%. This excess is consistent with additional surface-related background components not included in the simplified simulation, such as radon, cosmogenic surface activity in the cryostat, and inert gases released by the reactor.

\section{\label{sec:bkgmodel}Background model}

The background model for run 2 follows the same general approach used in run 1~\cite{Ackermann:2026vkq}: the individual background components -- including cosmogenic activation of the crystal and surrounding materials, muon-induced events, environmental and intrinsic radon, and residual gamma contamination -- are simulated and fitted simultaneously to the measured spectra of each detector. The simulations are performed in  Geant4 version 10.4.3 and use the MaGe framework~\cite{Boswell:2010mr}. Charge collection effects due to the detector geometry and Lindhard quenching~\cite{Bonhomme:2022lcz} are applied in a post processing step on an event-by-event basis. A detailed model of the \conusplus~setup including the new 2.42~kg detectors and the full shield structure is implemented. The simulation of the cosmic ray components (muons and neutrons), as well as the reactor neutrons contribution, are based on the results in \cite{CONUS:2024vyx}. The flux, energies, and angular distributions of muons in the \conusplus~room are calculated from models for shallow underground laboratories in literature \cite{Reyna:2006tf, Bugaev:1998ba}. Based on the overburden, a muon flux of  ($60.6 \pm 3.6$) s$^{-1}$ cm$^{-2}$ is calculated in the \conusplus~room.  For the cosmic neutron spectra in the room, appropriate neutron spectra were taken from \cite{Gordon:2004} and propagated through a model of the KKL reactor building, resulting in a flux of (0.9 $\pm$ 0.2)~cm$^{-2}$~d$^{-1}$. More details are provided in~\cite{CONUS:2024vyx}. For the evaluation of the radon background contribution, $^{214}$Pb and $^{214}$Bi isotopes are placed uniformly on surfaces inside the detector chamber of the \conusplus~shield. Inert gases released by the reactor like $^{134}$Xe and $^{3}$H are placed homogeneously in the full volume of the chamber. Residual gamma contaminations in the materials of the setup which are considered are: $^{60}$Co, $^{59}$Co, $^{57}$Co, $^{54}$Mn, $^{71}$Ge, $^{68}$Ge, $^{68}$Ga, $^{65}$Zn, $^{3}$H, $^{55}$Fe, $^{71m}$Ge, $^{73m}$Ge, and $^{75m}$Ge from cosmogenic activation in the germanium crystals, $^{60}$Co, $^{57}$Co, and $^{54}$Mn from cosmogenic activation in the copper parts of the cryostats, and $^{210}$Pb and its daughters in the inner part of the lead shield. \newline
As in run 1, the model is developed and validated using only data above 0.4~keV$_{ee}$, outside the blinded region, where the expected \CEvNS~signal is negligible, in accordance with the blind analysis strategy described in Sec.~\ref{sec:likelihood}. In particular, a combined fit of all of the components to the background only region in reactor on and off data from 0.4~keV$_{ee}$ to 28~keV$_{ee}$ is used to get priors for the scaling of the individual components in the likelihood fit. The extended high energy range up to 400~keV$_{ee}$ is used to get information on the intensity of the radon-induced gamma peaks as well as certain gamma peaks induced by metastable germanium states. \newline

As discussed in Sec.~\ref{sec:exp}, one of the original 1~kg detectors, C3, was deliberately kept in place through the run 2 upgrade rather than replaced to directly assess background stability. The background level and spectral shape observed for C3 in run 2 are consistent with those measured in run 1~\cite{Ackermann:2025obx}, and no new contamination -- in particular no additional $^{60}$Co activity from surface contamination -- was introduced into the shield during the upgrade~\cite{CONUS:2024vyx}.

\subsection{\label{sec:deadlayer}Dead and transition layer estimation}

An important input for the background simulations and the correct signal prediction is the thickness of the dead layer, the Li-diffused n+ layer around the germanium crystal with zero charge collection, and the transition layer (TL), the area between dead layer (DL) and bulk volume, where charge collection is suppressed. The sum of these two layers is called the partial charge collection layer (PCCL). A thicker PCCL leads to overall suppression of low energy signals in the crystal and a reduction of its total active volume. In particular, a larger transition layer shifts high energy events to lower energies due to the partial charge collection. In the simulations, the positions of all energy depositions are saved and based on their distance to the nearest crystal surface, the following model is used for the charge collection efficiency:

\begin{equation}
    \epsilon_{cc} = \frac{1}{e^{-(x - (DL + 0.5\,TL))/(0.17\,TL)} + 1}
\end{equation}

In this way, $\epsilon_{cc} \approx 0$ for $0 < x < DL$, $0 < \epsilon_{cc} < 1$ for $DL < x < PCCL$, and $\epsilon_{cc} \approx 1$ for $x > PCCL$. For the two 2.42~kg detectors in run 2, C6 and C7, the manufacturer specifies a PCCL thickness of $(0.6^{+0.2}_{-0.1})$~mm. The split between DL and TL was determined by comparing the shape of run 2 data without application of the muon veto to cosmic muon simulations and calculating the $\chi^2$/ndf difference between the two for different DL and TL values in the simulation post-processing, with their sum constrained to the manufacturer value. In this way, a DL thickness of about 0.25~mm and a TL thickness of about 0.35~mm were found to best describe the data for both detectors. For C3, a DL thickness of 0.48~mm and a TL thickness of 0.16~mm were previously determined with gamma attenuation measurements with a $^{241}$Am source.

\subsection{Background model for the 2.42~kg detectors}

Compared to the original 1~kg detectors, the new 2.42~kg detectors
achieve a reduction in background level per unit mass of approximately
a factor of two. This improvement arises from two effects: the selection of more radiopure materials for the cryostats of the new detectors, and the more favorable mass-to-surface ratio
of the larger crystals, which reduces the relative contribution of
surface-related backgrounds. Since the shielding and veto environment
shared by all detectors was shown to be unchanged by the upgrade, this
improvement can be attributed directly to the new detector hardware
rather than to any change in the surrounding setup.

\begin{figure}[htb]
    \centering
    \includegraphics[width=\columnwidth]{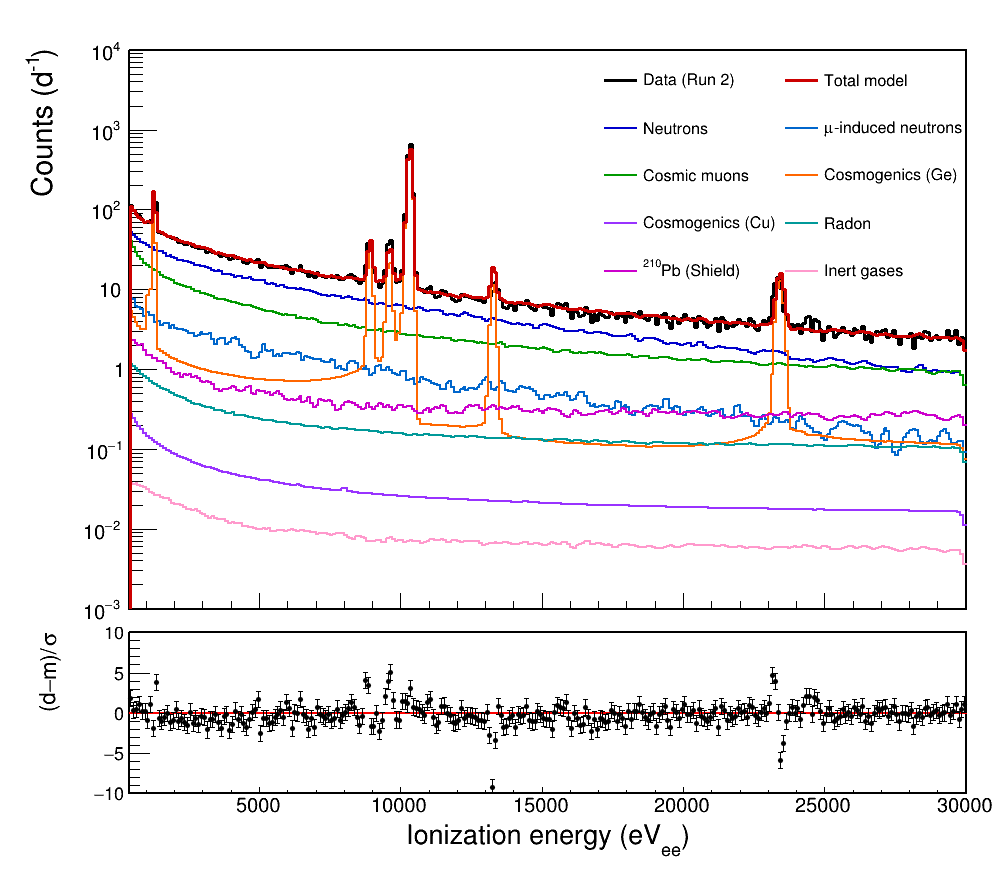}
    \caption{Background model for detector C7 compared to the energy spectrum measured
    between 0.4~keV$_{ee}$ and 30~keV$_{ee}$. The individual background components are shown separately.}
    \label{fig:bkgmodel_lowE}
\end{figure}

\begin{figure}[htb]
    \centering
    \includegraphics[width=\columnwidth]{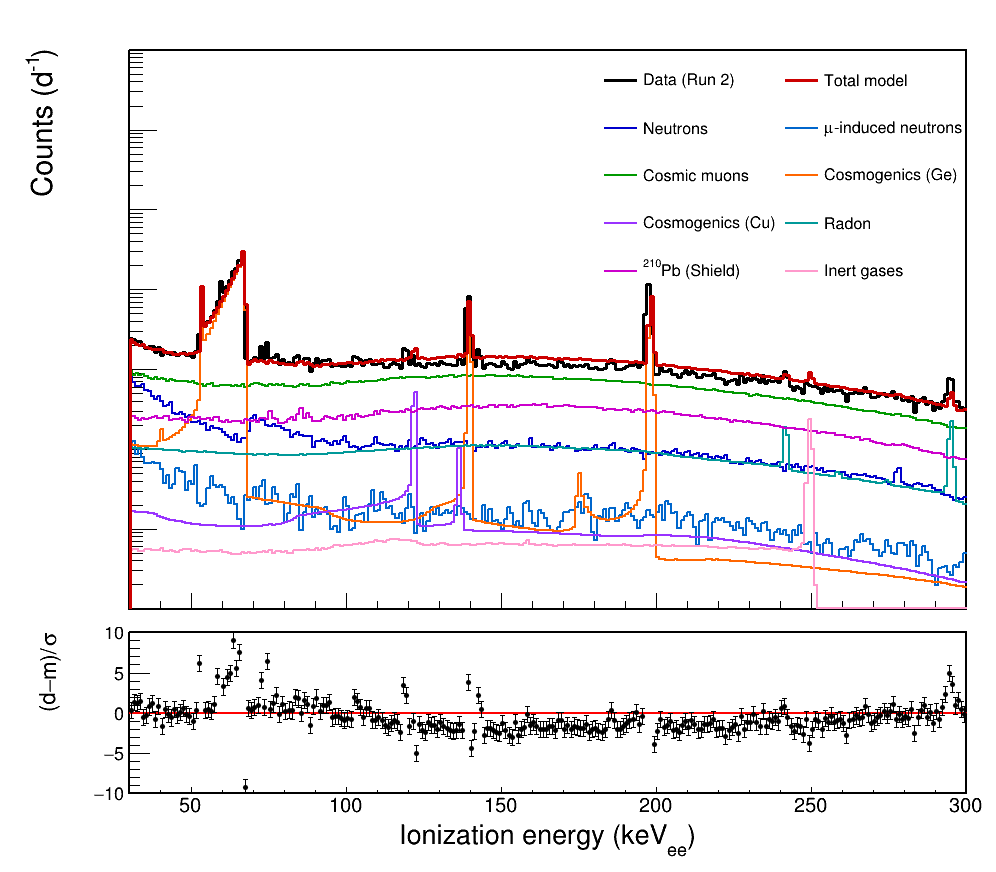}
    \caption{Background model for detector C7 compared to the energy spectrum measured
    between 30~keV$_{ee}$ and 300~keV$_{ee}$. The individual background components are shown separately.}
    \label{fig:bkgmodel_highE}
\end{figure}

Figures~\ref{fig:bkgmodel_lowE} and \ref{fig:bkgmodel_highE} show the resulting background model for detector C7, one of the new large detectors, compared to the measured spectrum, separately for the low-energy range below 30~keV$_{ee}$ and the high-energy range up to 300~keV$_{ee}$. The model reproduces the data across the full energy range and for all detectors, with the fitted contributions from neutrons, muons, radon, and crystal and copper cosmogenic activation shown individually. As in run 1, the background in the \CEvNS~ROI is dominated by cosmogenic components, mainly cosmic-ray neutrons and muon-induced events. At energies above 100~keV, the spectrum is still dominated by muon-induced events, while the influence of cosmic neutrons drops off in comparison. Apart from muons, $^{210}$Pb from the inner shield layers and radon in the detector chamber are the prominent sources of background in this energy range.   Equivalent figures for the C3 and C6 detectors can be found in Section~\ref{sec:appendix}. 

The stability of the background rate throughout run 2 was monitored for all detectors in three energy regions: the \CEvNS~ROI, the 2--8~keV$_{ee}$ range, and the region above 100~keV$_{ee}$. Fig.~\ref{fig:stability_bkg} shows this monitoring for the individual detectors. A good stability is observed across the whole data-taking period in all three regions supporting the assumption of a time-independent background model over the course of the run.

\subsection{\label{sec:bkg_difference}Difference in reactor on and reactor off model}

As described in Section \ref{sec:sensitivity}, the likelihood fit which yields the final CE$\nu$NS result, fits both the reactor on and off data measured in run 2 of the experiment. For this reason, a separate background model for the reactor off data is developed. The main difference to the reactor on model is the flux of cosmic neutrons and muons in the room. During reactor maintenance the drywell lid of the reactor pressure vessel is placed directly above the \conusplus~experimental room, leading to an increase of approximately 0.25~m w.e. of overburden above the experiment. As previously validated in run 1 of the experiment, this leads to a decrease of the cosmic muon flux in the room by 3$\%$ and a decrease of the cosmic neutron flux by 19$\%$. The corresponding contributions to the background model are also reduced accordingly during this time. Compared to the model in run 1, which showed variations of the radon contribution during the data-taking period, the run 2 model features almost constant radon contributions, in particular between on and off phases. This is due to an improved performance of the radon flushing system which was achieved through a new set of tubes and pressure reducers at the radon-free air bottles. 

\section{\label{sec:prediction}Signal prediction}

The expected \CEvNS~signal for run 2 is calculated using the same general framework as in run 1 \cite{Ackermann:2025obx}, updated with the reactor and detector parameters specific to this dataset. The \conusplus~detectors are positioned at a distance of 20.7~m from the reactor core center, as described in Sec.~\ref{sec:exp}. 

The reactor thermal power and fuel composition are taken from official reactor operation documents for the run 2 period. The average thermal power is 3552~MW, where a conservative relative uncertainty of 1.5$\%$ is assumed. The reactor antineutrino flux depends on the fission fractions of the four dominant fissile isotopes, $^{235}$U, $^{238}$U, $^{239}$Pu, and $^{241}$Pu, which for the run 2 reactor-on period average to 0.54, 0.08, 0.31, and 0.07, respectively.

For the antineutrino energy spectrum, we follow a data-driven approach based on the Daya Bay measurement~\cite{DayaBay:2021dqj}, adjusted to the fission fractions of the KKL reactor. This baseline spectrum is further refined above 8~MeV using a dedicated Daya Bay measurement~\cite{DayaBay:2022eyy}, and supplemented below the inverse-beta-decay (IBD) threshold of 1.8~MeV with updated summation calculations~\cite{Estienne:2019ujo}, following the same approach as in previous \conusplus~analyses~\cite{Ackermann:2024kxo,Ackermann:2025obx}. The nuclear recoil spectrum is converted to the observed ionization energy using the Lindhard quenching model, with a quenching factor $k = 0.162 \pm 0.004$, as previously measured in~\cite{Bonhomme:2022lcz}. The predicted signal further incorporates the detector-specific active mass, trigger efficiency, energy resolution, and non-linearity correction described in Sec.~\ref{sec:det}.

The resulting expected number of \CEvNS~signal events for the run 2 exposure, integrated over the ROI from the energy threshold of each detector up to 800~eV$_{ee}$, is given in Table~\ref{tab:signalpred} for each detector individually and for the combination of the three, yielding a total of $1701 \pm 198$ predicted signal events. Given a threshold between 140 and 170~eV$_{ee}$ depending on the detector, the predicted signal is sensitive to reactor antineutrinos with energies down to 5~MeV.

\begin{table}[htb]
    \centering
    \caption{Predicted number of \CEvNS~signal events for the run 2
    exposure, per detector and combined.}
    \label{tab:signalpred}
    \renewcommand{\arraystretch}{1.7}
    \begin{tabular}{lc}
        \hline\hline
        Detector & Predicted signal events \\
        \hline
        C3 & $195 \pm 23$ \\
        C6 & $604 \pm 73$ \\
        C7 & $903 \pm 111$ \\
        \hline
        Combined & $1701 \pm 198$ \\
        \hline\hline
    \end{tabular}
\end{table}

\begin{figure}[htb]
    \centering
    \includegraphics[width=\columnwidth]{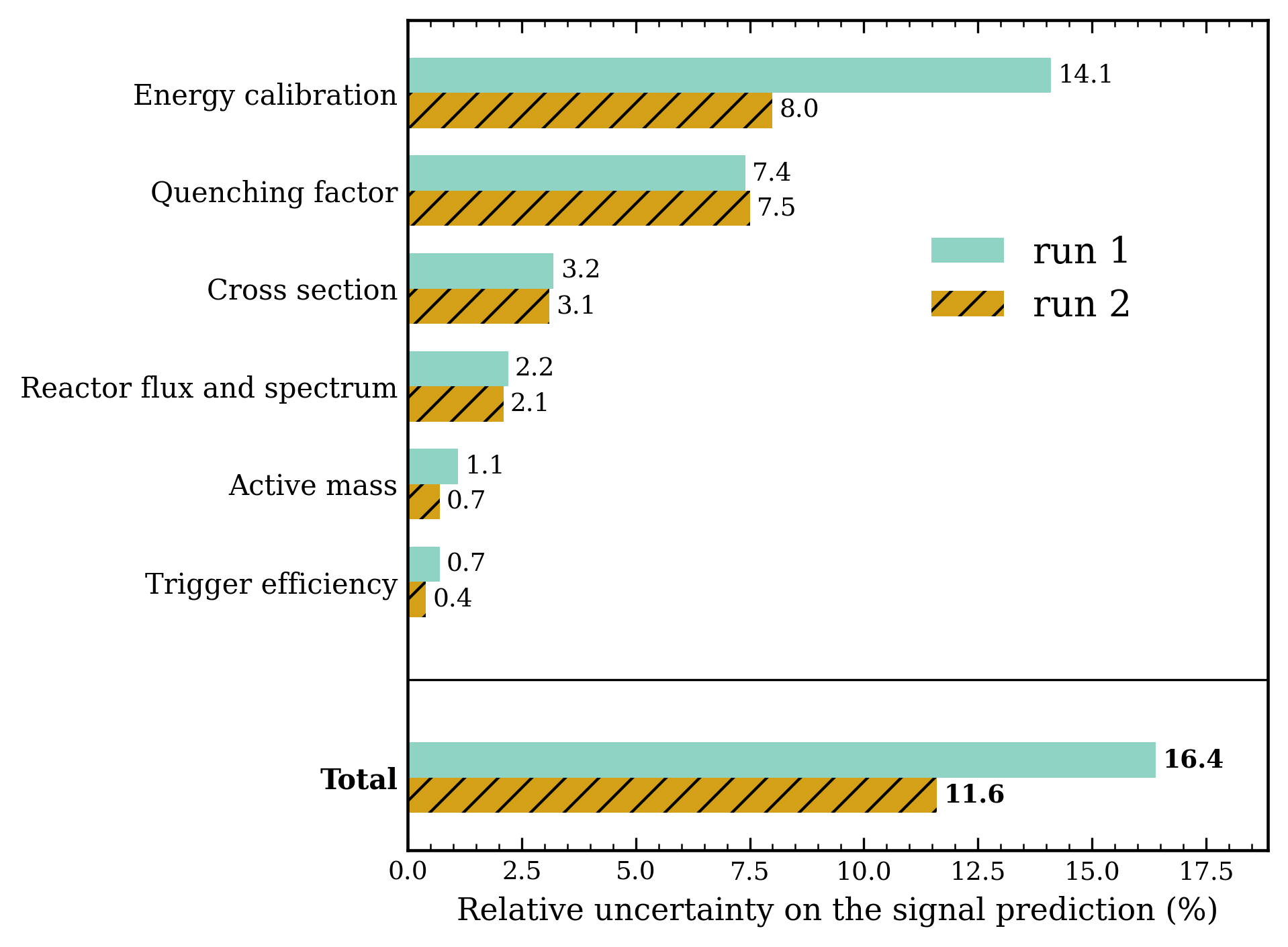}
    \caption{Breakdown of the systematic contributions to the uncertainty on the predicted \CEvNS~signal, compared between run 1 and run 2.}
    \label{fig:sysbreakdown}
\end{figure}

The uncertainty on the combined prediction is broken down by source in Fig.~\ref{fig:sysbreakdown} compared to the corresponding run 1 values \cite{Ackermann:2025obx}. Thanks to the improved energy-scale calibration of 3~eV$_{ee}$ described in Sec.~\ref{sec:det}, the uncertainty on the predicted signal arising from the energy threshold is reduced from 14.1$\%$ in run 1 to 8.0$\%$ in run 2, becoming similar to the quenching-factor uncertainty, which remains essentially unchanged at 7.4--7.5\%. The \CEvNS~cross-section uncertainty (3.1--3.2\%) and the reactor-flux (2.1--2.2\%) are also among the leading contributions, while the active-mass determination and the trigger-efficiency uncertainty are each at or below the one-percent level. The total uncertainty on the signal prediction is reduced from 16.4$\%$ in run 1 to 11.6$\%$ in run 2. A dedicated future irradiation campaign with a $^{252}$Cf source, would further reduce the energy-scale uncertainty below 2~eV$_{ee}$, corresponding to an energy threshold uncertainty contribution of about 4\%~\cite{SanchezGarcia:2026nvr}, thereby making it subdominant with respect to the quenching-factor and reactor-flux uncertainties and reducing the total signal-prediction down to 9.3$\%$. 

\section{Statistical framework}
\label{sec:likelihood}

The signal is extracted through a binned likelihood fit of the reactor-on and reactor-off energy spectra of the three run 2 detectors C6, C7, and C3, performed simultaneously. The framework builds on the run 1 analysis~\cite{Ackermann:2025obx}, but differs from it in three main points. First, the uncertainty on the background model, which in run 1 was evaluated outside the fit and added in quadrature, is now included in the likelihood through nuisance parameters. Second, the single background scaling factor of run 1 is replaced by independent normalizations for each background component, constrained by an extended fit range. Third, the likelihood is parametrized directly in terms of the number of signal counts, instead of a signal strength relative to the prediction, so that the correlations between the signal and the nuisance parameters are propagated to the result.

As in run 1, two independent likelihood implementations, A and B, were developed. Both follow the formalism described below, but differ in the treatment of the energy resolution and of some systematic uncertainties. In particular, the energy-scale non-linearity is included as nuisance parameters only in Likelihood B. The overall likelihood is constructed as
\begin{align}
    -2\log\mathcal{L} ={}& -2\log\mathcal{L}_{\mathrm{ON}}
    -2\log\mathcal{L}_{\mathrm{OFF}} \nonumber\\
    & + \sum_{i,j=1}^{N_{\mathrm{c}}}(\theta_i-\bar\theta_i)\,
    \left(\mathrm{Cov}^{-1}\right)_{ij}\,(\theta_j-\bar\theta_j) \nonumber\\
    & + \sum_{l=1}^{N_{\mathrm{u}}} \frac{(\theta_l-\bar\theta_l)^2}{\sigma_{\theta_l}^2},
    \label{eq:likelihood}
\end{align}
where $\mathcal{L}_{\mathrm{ON}}$ and $\mathcal{L}_{\mathrm{OFF}}$ are the binned Poisson likelihoods for the reactor-on and reactor-off periods, and the remaining terms are Gaussian pull terms constraining the nuisance parameters $\theta$ to their central values $\bar\theta$. The $N_{\mathrm{c}}$ correlated parameters, such as the trigger efficiency and calibration parameters of Sec.~\ref{sec:det}, are constrained through their covariance matrix $\mathrm{Cov}$. The $N_{\mathrm{u}}$ uncorrelated parameters, such as the active mass, the energy-scale uncertainty and the background normalizations, are constrained independently. The binned Poisson likelihood for a single period has the form
\begin{equation}
    -\log\mathcal{L} = \sum_{i=1}^{N}
    \left[-n_i\log(\mu_i) + \mu_i + \log(n_i!)\right],
    \label{eq:poisson}
\end{equation}
where $N$ is the number of bins in the ROI, $n_i$ is the measured bin content, and $\mu_i$ is the corresponding model prediction. For each detector $d$, the reactor-on and reactor-off model predictions are
\begin{align}
    \mu_i^{\mathrm{ON},d} ={}& N_s\,w_d\,f_i^{s,d}(\vec\eta)
    + t_{\mathrm{ON}}\sum_{k=1}^{7} b_k\,n_{i,k}^{b}\,
    c_{\mathrm{dt}}\,\frac{\theta_{\mathrm{det}}}{m_{\mathrm{act}}},
    \label{eq:muon}\\
    \mu_i^{\mathrm{OFF},d} ={}& \; t_{\mathrm{OFF}}\sum_{k=1}^{7} b_k\,n_{i,k}^{b}\,
    c_{\mathrm{dt}}\,\frac{\theta_{\mathrm{det}}}{m_{\mathrm{act}}},
    \label{eq:muoff}
\end{align}
where $N_s$ is the number of \CEvNS~counts in the ROI, summed over the three detectors, which is the parameter of interest. $f_i^{s,d}(\vec\eta)$ is the predicted signal spectral shape of detector $d$, normalized to unity in the ROI, and $w_d$ is the expected fraction of the signal counts in detector $d$, given by its exposure, dead time and detection efficiency. In Likelihood B, the signal shape is corrected for the energy-scale non-linearity, modeled as a second-order polynomial in energy with coefficients $\vec\eta = (\eta_0,\eta_1,\eta_2)$, as described in Sec.~\ref{sec:calibration}. $n_{i,k}^{b}$ is the predicted spectrum of background component $k$, and $b_k$ its normalization, constrained via a pull term to the value obtained from the background-model fit of Sec.~\ref{sec:bkgmodel}. The seven background components are the muon-induced, neutron-induced, cosmogenic, radon, reactor-released gas, $^{210}$Pb, and copper-cryostat backgrounds described in Sec.~\ref{sec:bkgmodel}. $t_{\mathrm{ON}}$ and $t_{\mathrm{OFF}}$ are the reactor-on and reactor-off live times, $c_{\mathrm{dt}}$ is the dead-time correction factor accounting for the combined TRP-reset and muon-veto dead time described in Sec.~\ref{sec:analysis} and the DAQ dead time, $m_{\mathrm{act}}$ is the detector active mass, and $\theta_{\mathrm{det}}$ summarizes uncertainties on the detector response. The background normalizations $b_k$ and, in Likelihood B, the non-linearity coefficients $\vec\eta$ are treated as nuisance parameters in the fit, each constrained by a pull term to its central value and uncertainty as determined in Secs.~\ref{sec:bkgmodel} and~\ref{sec:calibration}, respectively. In run 1, both contributions were instead estimated by varying the background model and the non-linearity correction outside the fit~\cite{Ackermann:2025obx}.

The ROI used in the fit runs from the energy threshold of each detector up to 800~eV$_{ee}$. In contrast to run 1, where only the ROI was fitted, a dedicated background-only region from 0.8 to 28~keV$_{ee}$ is included to constrain the seven background normalizations $b_k$, while the spectral region up to 400~keV$_{ee}$ is used to additionally constrain the radon contribution. The background normalizations $b_k$ are shared between the reactor-on and reactor-off periods, and differences in both periods stemming from the larger overburden during reactor-off periods are accounted for in the muon and neutron spectra. The statistical uncertainty of the simulated background spectra is included in the fit via a Barlow–Beeston-lite treatment~\cite{Barlow:1993dm}. In each bin, the combined Monte Carlo variance of the background components (scaling as the square of each component's fitted normalization) is absorbed by treating the bin's predicted mean as Gamma-distributed rather than fixed. Marginalizing the Poisson likelihood over this Gamma-distributed mean yields a closed-form Negative Binomial likelihood, adding no free parameters and reducing to the ordinary Poisson term as the Monte Carlo uncertainty vanishes. 

Likelihood A is used as default. The contribution of the non-linearity, only implemented in Likelihood B, is added in quadrature, and the difference between both implementations is included as an additional systematic uncertainty, as discussed in Sec.~\ref{sec:result}.

Since $N_s$ is directly the parameter of interest, the best-fit value and uncertainty of the fit are taken as the final result. Throughout this paper, the strength of the signal is quantified by $N_s/\sigma(N_s)$, the inverse of the relative precision $\sigma(N_s)/N_s$, where $\sigma(N_s)$ is the total uncertainty on the fitted number of signal counts. Uncertainties on the signal prediction, such as the reactor antineutrino flux and the quenching factor normalization, do not enter the fit, but only the comparison of $N_s$ to the SM expectation. As in run 1, all prediction uncertainties described in Sec.~\ref{sec:prediction} are conservatively treated as fully correlated between detectors in this comparison. As a cross-check, the fit was also performed in terms of the signal strength, with the result converted to the number of counts by Monte Carlo sampling of the fit parameters from their covariance matrix, giving a result consistent with the direct fit.

The covariance matrix of the fit is also used to split the total uncertainty on $N_s$ into its individual contributions. For a group $G$ of nuisance parameters (for example the trigger efficiency or the background normalizations), the remaining variance of $N_s$ when the parameters of $G$ are fixed is
\begin{equation}
    \sigma^2_{N_s|G} = \Sigma_{NN} - \Sigma_{NG}\,\Sigma_{GG}^{-1}\,\Sigma_{GN},
    \label{eq:condvar}
\end{equation}
where $\Sigma$ is the covariance matrix of the fit parameters. The contribution of $G$ to the total uncertainty is then given by the difference in quadrature $\sigma^2_G = \Sigma_{NN} - \sigma^2_{N_s|G}$. Fixing all nuisance parameters at once leaves only the statistical uncertainty, given by $\sigma^2_{\mathrm{stat}} = 1/(\Sigma^{-1})_{NN}$.

We followed a blind analysis strategy. The measured spectra in the signal ROI, below 400~eV$_{ee}$, remained hidden until the analysis framework had been validated. We adopted a staged unblinding protocol, in which the results of Likelihood A and B were compared at each step before proceeding to the next one. The likelihood modifications introduced for run 2 were first validated on the published run 1 dataset, yielding $N_s = 399 \pm 103$ counts with a significance of 3.9$\,\sigma$, compared to 3.7$\,\sigma$ in the original analysis~\cite{Ackermann:2025obx}. The improvement is small, since the run 1 result is dominated by the statistical uncertainty. The modifications were then tested on Asimov datasets constructed for the run 2 exposure with injected signals, before any unblinding of real data took place. The detectors were then unblinded one by one, starting with C3, whose behaviour has already been known from run 1, followed by C7 and C6. The stability of the signal was checked for extended fit ranges. The Asimov datasets used to validate the likelihood are also used to project the expected sensitivity of the run 2 analysis, as discussed in the next section.

During the unblinding, several changes were made, each motivated by checks independent of the signal. When unblinding C3, a deficit of events below 300~eV$_{ee}$ with respect to the run 1 spectrum revealed an error in the implementation of one of the waveform quality cuts. After its correction, the deficit disappeared. Flat waveforms from delayed events in accidental coincidence, which are not removed by the quality cuts, were included as an additional contribution to the background model. Two independent models of this contribution agree in shape and differ by less than a factor of two in normalization. Additionally, the reactor-off spectra of all three detectors showed disagreement to the background model below 5~keV. This prompted an investigation of the application of the muon veto rejection in the simulations and the veto system was fully included in the Monte Carlo studies, which resolved the issue. Finally, the background normalizations were shared between the reactor-on and reactor-off periods, and the statistical uncertainty of the simulated background spectra was included, which reduces the significance by about 10\%. After unblinding, following internal review, the manufacturer value of $(0.6^{+0.2}_{-0.1})$\,mm for C6 and C7, originally interpreted as the DL thickness, was found to refer to the full PCCL. The background model was therefore re-evaluated with the DL and TL thicknesses constrained to this PCCL value (Sec.~\ref{sec:deadlayer}). The previous model yields $N_s = 1555 \pm 152$, in agreement with the final result (Table~\ref{tab:crosschecks}).

\section{\label{sec:sensitivity}Sensitivity}

Using the likelihood framework described above, together with the run 2 exposure and signal prediction described in Secs.~\ref{sec:datataking} and \ref{sec:prediction}, the expected sensitivity is evaluated on Asimov datasets as a function of exposure.

\begin{figure}[htb]
    \centering
    \includegraphics[width=\columnwidth]{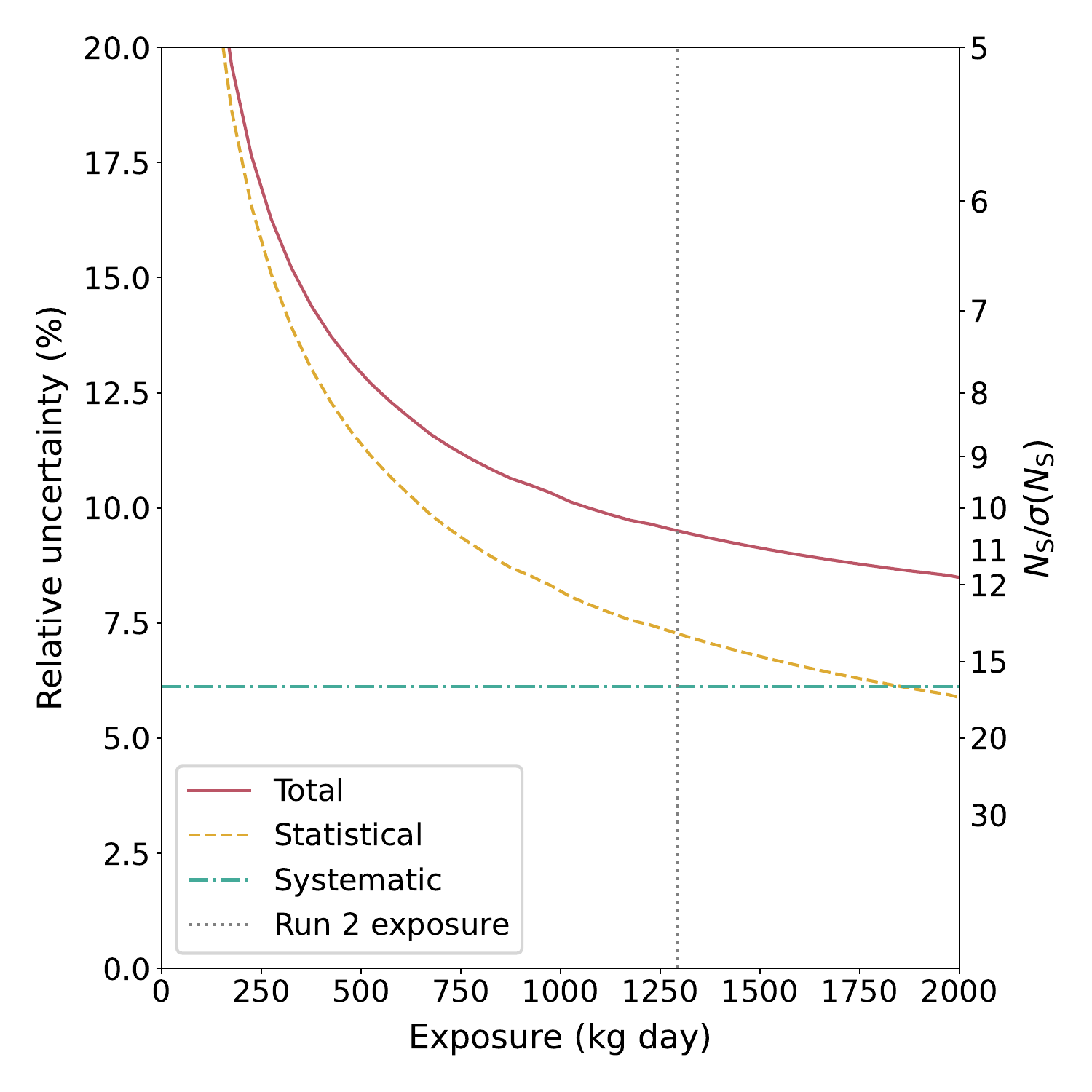}
    \caption{Expected $N_s/\sigma(N_s)$ (right axis) for the SM \CEvNS~signal, and relative uncertainty on the number of signal counts (left axis), as a function of exposure, for the run 2 dataset. The total, statistical and systematic uncertainties are shown as solid, dashed and dash-dotted lines, respectively. Run 2 exposure is indicated with dotted vertical black line.}
    \label{fig:sensexposure}
\end{figure}

Figure~\ref{fig:sensexposure} shows the expected value of $N_s/\sigma(N_s)$ for the SM \CEvNS~signal, together with the relative uncertainty on $N_s$, as a function of exposure. The statistical uncertainty decreases with exposure, while the systematic uncertainty remains approximately constant, so that both become of comparable size at the full run 2 exposure of 1296~kg$\cdot$days reactor-on. The expected value of $N_s/\sigma(N_s)$ is 10.5, with a total relative uncertainty on $N_s$ of 9.5\%, of which 7.3\% is statistical and 6.1\% systematic. The analysis therefore remains statistics-dominated, as in run 1.

\begin{figure}[htb]
    \centering
    \includegraphics[width=\columnwidth]{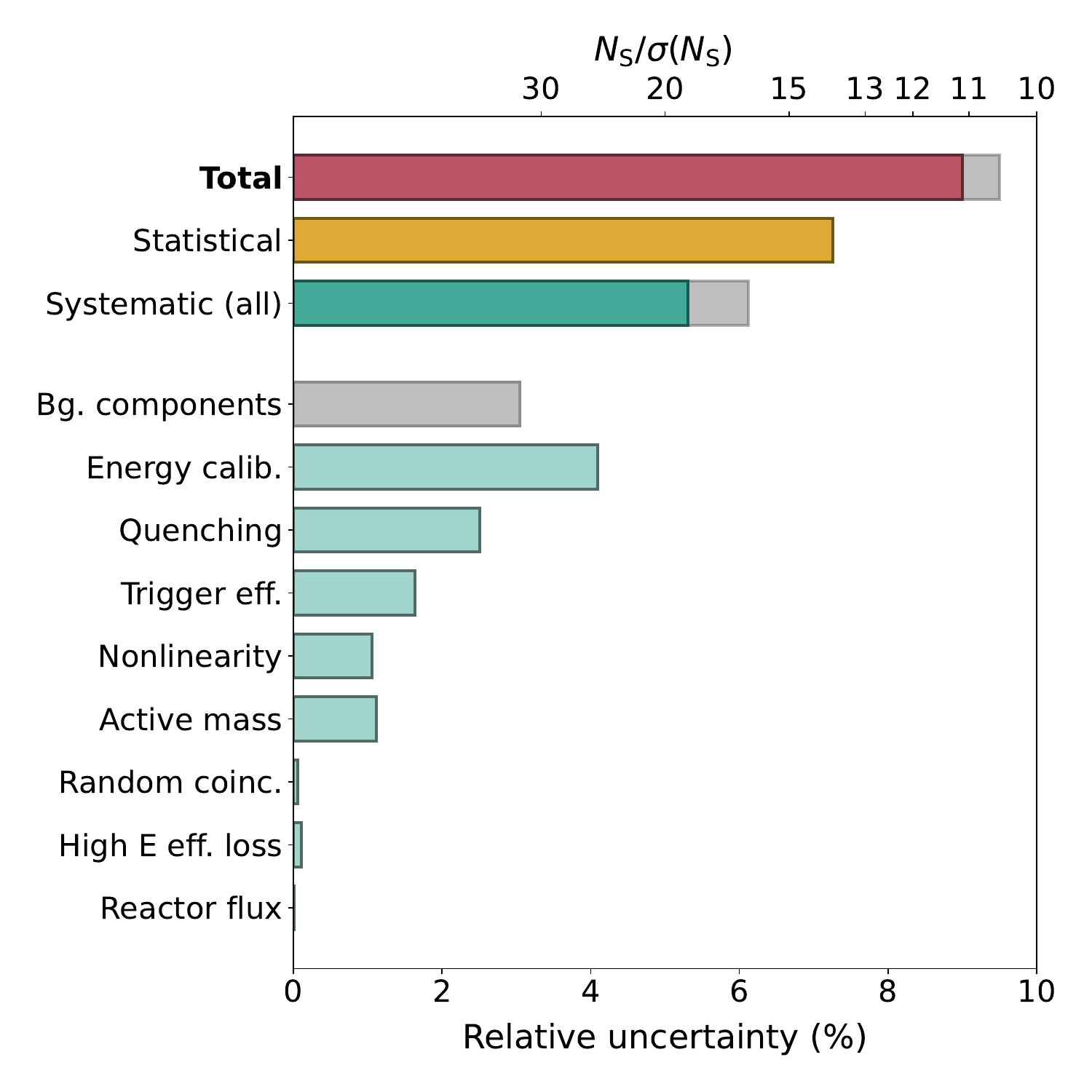}
    \caption{Breakdown of the individual contributions to the expected relative uncertainty on the number of signal counts (bottom axis) and the corresponding $N_s/\sigma(N_s)$ (top axis), at the full run 2 exposure. From top to bottom: total uncertainty, statistical uncertainty, combined systematic uncertainty, and its individual components. The background-model contribution (grey) is taken from the fit to the run 2 data and added in quadrature. The grey extensions of the total and systematic bars show its impact.}
    \label{fig:sysbreakdown2}
\end{figure}

Figure~\ref{fig:sysbreakdown2} shows the breakdown of the individual systematic contributions to the expected uncertainty on $N_s$ at the full run 2 exposure, evaluated with Eq.~(\ref{eq:condvar}), together with the corresponding value of $N_s/\sigma(N_s)$. The background-model contribution is not evaluated on the Asimov datasets. It is taken from the systematic breakdown of the likelihood fit to the run 2 data (Sec.~\ref{sec:result}), where the normalizations $b_k$ of the seven background components are treated as nuisance parameters, and added in quadrature to the other contributions. The largest contributions come from the energy calibration (4.1\%) and the background model (3.3\%), followed by the quenching factor (2.3\%) and the trigger efficiency (1.6\%). The non-linearity and the active mass contribute about 1\% each, while random coincidences and the residual detection-efficiency loss at high energy are negligible. Since the result is expressed as a number of counts, the quenching factor affects $N_s$ through its impact on the shape of the signal spectrum and on the background prediction, while its impact on the signal normalization, together with the reactor antineutrino flux, enters when $N_s$ is compared to the SM expectation.

In the comparison with the SM expectation, the dominant uncertainties on the signal prediction are the energy calibration at the threshold position (8.0\%) and the quenching factor (7.5\%). Both act in the same way. The signal is concentrated close to the energy threshold, where the recoil spectrum falls steeply. A small change of the threshold position or of the quenching factor shifts the predicted ionization spectrum with respect to the threshold, and therefore changes the expected number of counts above threshold by a large amount. Including the full systematic budget on the signal prediction of Sec.~\ref{sec:prediction} (11.6\%), the expected total relative uncertainty on the measured cross section with respect to the SM prediction is approximately 15\%. This comparison is therefore limited by the signal prediction rather than by the measurement, and reaches a precision comparable to that of the most recent COHERENT germanium measurement of \CEvNS~\cite{COHERENT:2026yje}.

The main systematic contributions can be reduced in the future. The energy-calibration uncertainty at the threshold position can be reduced below 2~eV$_{ee}$ with neutron irradiation campaigns, as demonstrated in~\cite{SanchezGarcia:2026nvr}. This also reduces the correlated trigger-efficiency uncertainty, and lowers the energy-calibration contribution to the signal prediction to about 4\%, and the total uncertainty on the prediction to about 9\% (Sec.~\ref{sec:prediction}). The background-model uncertainty is dominated by the cosmic-ray neutron component~\cite{CONUS:2024vyx, Bonet:2021wjw}, which can be constrained with additional background measurement campaigns at KKL. New quenching-factor measurements are planned to reduce the second leading uncertainty on the signal prediction. Finally, the active-mass uncertainty can be reduced with dedicated measurements at MPIK.

\section{\label{sec:result} run 2 \CEvNS~results}

The results presented in this section are based on the run 2 dataset described in Sec.~\ref{sec:datataking}, collected with the detectors C6, C7, and C3. After all selection cuts, it corresponds to an exposure of 1296~kg$\cdot$days reactor-on and 86~kg$\cdot$days reactor-off. The analysis uses the energy thresholds of Sec.~\ref{sec:det} and the final analysis chain, including the changes introduced during the unblinding described in Sec.~\ref{sec:likelihood}.

\begin{figure*}[htb!]
    \centering
    \includegraphics[width=1.0\textwidth]{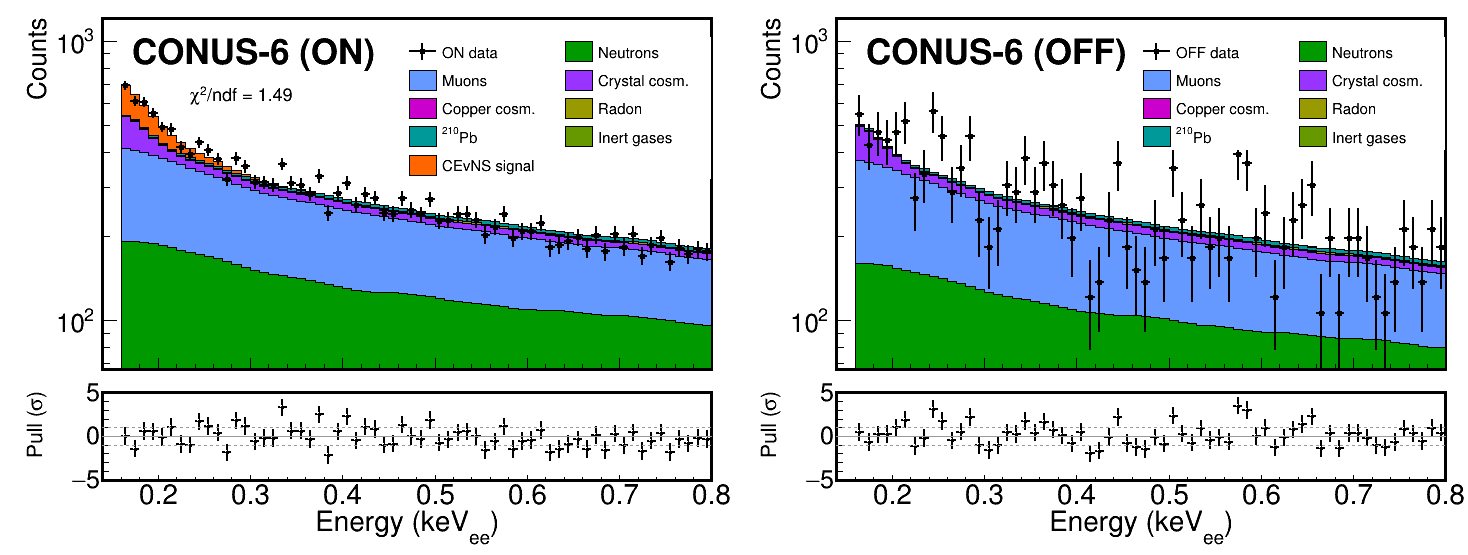}\\
    \includegraphics[width=1.0\textwidth]{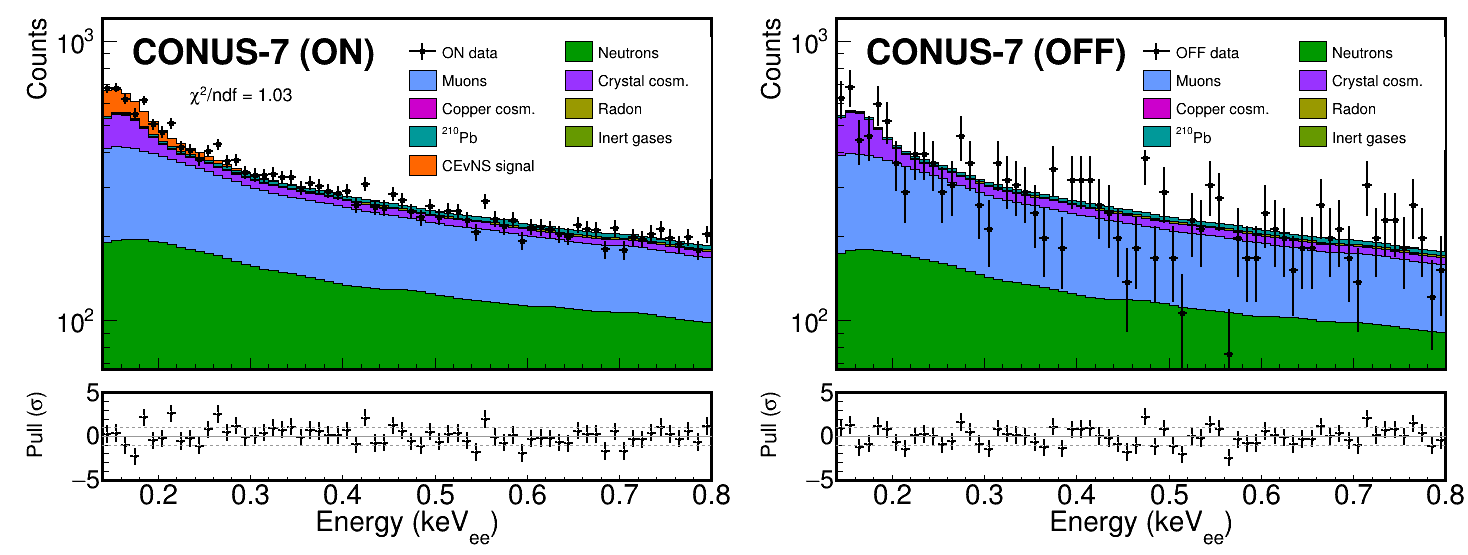}\\
    \includegraphics[width=1.0\textwidth]{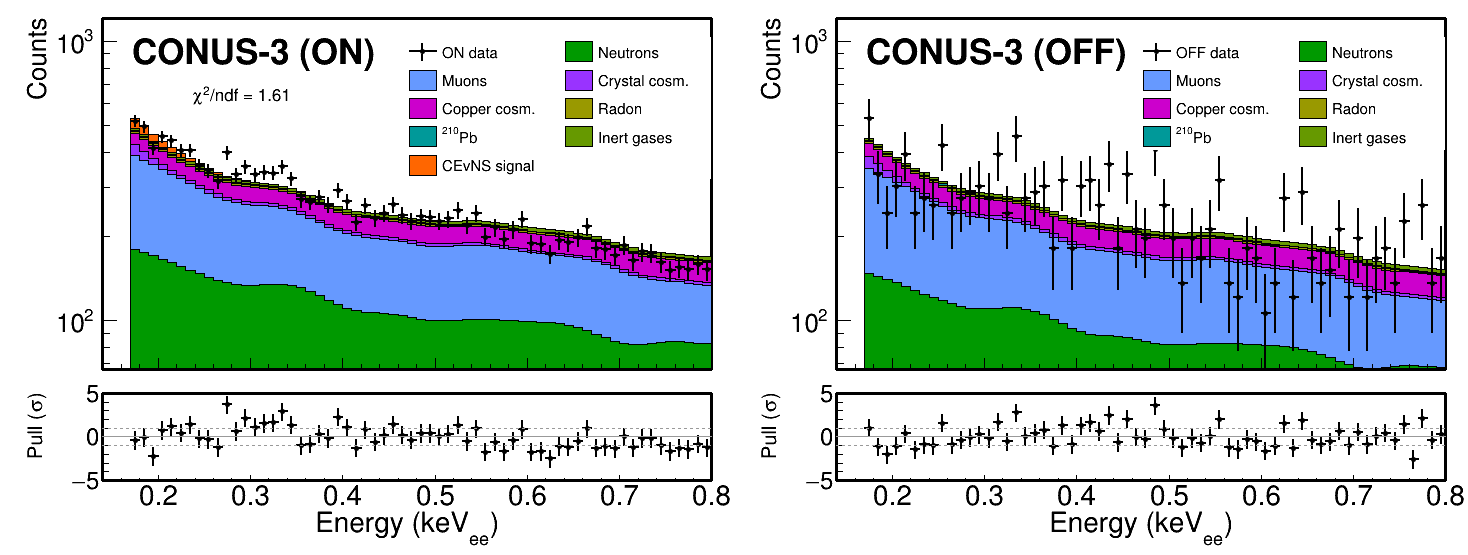}
    \caption{Reactor-on (left) and reactor-off (right) spectra of C6, C7, and C3 in the ROI, with the best-fit model. The stacked contributions are the background components and the \CEvNS~signal (orange). The lower panels show the pulls between data and model.}
    \label{fig:fit_individual}
\end{figure*}

Each detector is first fitted individually, using the same likelihood framework and nuisance parameters as in the combined fit. Figure~\ref{fig:fit_individual} shows the reactor-on (left) and reactor-off (right) spectra of C6, C7, and C3 in the ROI, together with the best-fit signal and background contributions. The lower panels show the residuals between data and model, normalized to the uncertainty of each bin. The \CEvNS~signal is visible in the reactor-on spectra as an excess close to the energy threshold. The $\chi^2$/ndf values of the individual fits, including both the reactor-on and reactor-off spectra, are 1.49 (C6), 1.03 (C7), and 1.61 (C3), for about 110 degrees of freedom. Since the asymptotic $\chi^2$ distribution is not guaranteed to hold for bins with few counts, as in the reactor-off spectra, the p-values are evaluated with pseudo-experiments generated from the best-fit model and fitted in the same way as the data, giving $2\times10^{-4}$ (C6), 0.37 (C7), and $5.0\times10^{-5}$ (C3). For C6 and C3, the deviations are mostly distributed over the full ROI, with the largest local excess of the data over the model in C3 between about 270 and 370\,eV$_{ee}$. Evaluating the reactor-on contribution separately yields p-values of 0.02 (C6), 0.30 (C7), and $2\times10^{-3}$ (C3). For the reactor-off contribution, the p-values are $1.7\times10^{-3}$ (C6), 0.46 (C7), and $4.1\times10^{-3}$ (C3). For C6, the deviations thus arise mainly from the reactor-off spectrum, while for C3 they are present in both. C7, which has the largest expected signal, is well described, while C3 contributes only about 11\% of the expected signal. The impact of these deviations on the result is evaluated with the consistency checks of Table~\ref{tab:crosschecks}, discussed below, which show no indication of a bias of the extracted signal. The results of the individual fits are summarized in Table~\ref{tab:fit_results}. The signal is observed with $N_s/\sigma(N_s)$ of 6.5 in C6 and 8.0 in C7, while C3 reaches 2.4, owing to its smaller mass and higher threshold. All individual results agree with the SM expectation within 1.1 standard deviations, with C6 lying 0.8$\,\sigma$ above and C7 1.1$\,\sigma$ below the expectation, while C3 is in excellent agreement.

The combined fit of the three detectors yields $N_s = 1565 \pm 117\,(\mathrm{stat}) \pm 102\,(\mathrm{syst})$ signal counts, corresponding to a relative precision of 9.9\% ($N_s/\sigma(N_s) = 10.1$). The corresponding spectra are shown in~\cite{ConusPlusRun2PRL}. The fit yields $\chi^2$/ndf = 1.39 for 381 effective degrees of freedom, corresponding to a p-value of $3\times10^{-6}$, driven by the deviations in C6 and C3. The total relative uncertainty of 9.9\% is composed of a statistical uncertainty of 7.5\% and a systematic uncertainty of 6.5\%, so that the result remains statistics-dominated, as expected. The measured value of $N_s/\sigma(N_s)$ is in agreement with the expected value of 10.5 derived in Sec.~\ref{sec:sensitivity}, being slightly lower since the measured number of signal counts is below the SM expectation. The final dataset was analysed with both likelihood implementations. Likelihood B yields $N_s = 1575 \pm 150$, in agreement with the default Likelihood A within 0.1$\,\sigma$, despite the different treatment of the energy resolution and of some systematic uncertainties. The difference of 0.6\% between both results is included as an additional systematic uncertainty. The measured number of signal counts is in agreement with the SM expectation of $N_{\mathrm{SM}} = 1701 \pm 198$ counts, obtained with the signal prediction of Sec.~VII, corresponding to a ratio $N_s/N_{\mathrm{SM}} = 0.92 \pm 0.14$, where both uncertainties are combined in quadrature, neglecting their correlations.

\begin{table}[htb]
    \centering
    \caption{Number of signal counts $N_s$ obtained from the individual fits of each detector and from the combined fit, compared to the SM expectation $N_{\mathrm{SM}}$, together with $N_s/\sigma(N_s)$.}
    \label{tab:fit_results}
    \setlength{\tabcolsep}{6pt}
    \renewcommand{\arraystretch}{1.7}
    \begin{tabular}{lccc}
        \hline\hline
        Fit & $N_s$ & $N_{\mathrm{SM}}$ & $N_s/\sigma(N_s)$ \\
        \hline
        C6 & 707 $\pm$ 108 & 604 $\pm$ 73 & 6.5 \\
        C7 & 743 $\pm$ 93 & 903 $\pm$ 111 & 8.0 \\
        C3 & 191 $\pm$ 81 & 195 $\pm$ 23 & 2.4 \\
        \hline
        Combined & 1565 $\pm$ 155 & 1701 $\pm$ 198 & 10.1 \\
        \hline\hline
    \end{tabular}
\end{table}

\begin{figure}[htb]
    \centering
    \includegraphics[width=\columnwidth]{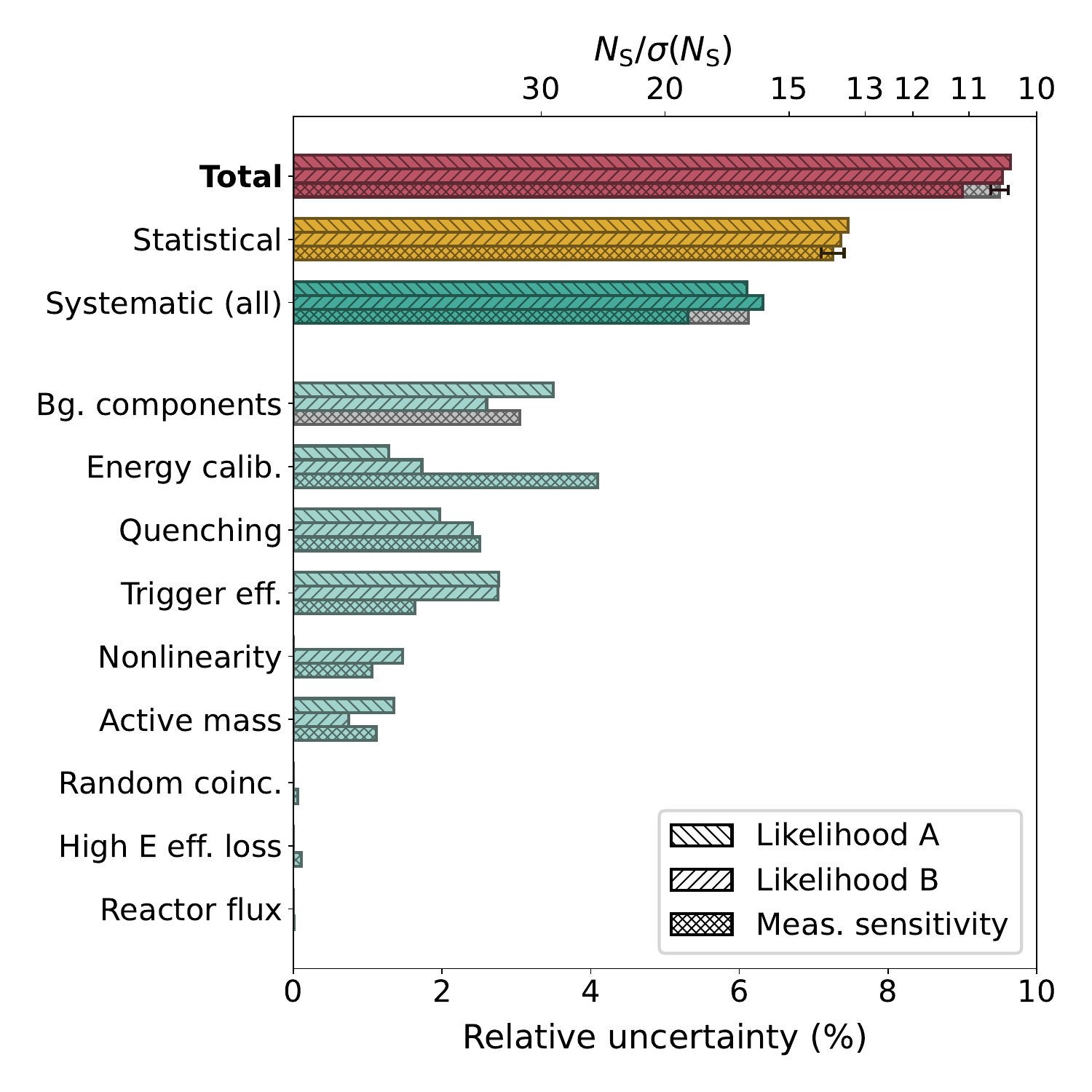}
    \caption{Breakdown of the relative uncertainty on the number of \CEvNS~events $N_s$ from the combined fit (bottom axis) and the corresponding $N_s/\sigma(N_s)$ (top axis), for Likelihood A and B, compared to the expected contributions from the sensitivity study (Sec.~\ref{sec:sensitivity}). From top to bottom, the total uncertainty, the statistical uncertainty, the combined systematic uncertainty and its individual components. The background-component contribution of the sensitivity study is taken as the average of the Likelihood A and B fits to the data.}
    \label{fig:sys_result}
\end{figure}

The individual contributions to the uncertainty on $N_s$ are evaluated from the covariance matrix of the combined fit using Eq.~(\ref{eq:condvar}). They are shown in Fig.~\ref{fig:sys_result} for both likelihood implementations, together with the expected contributions from the sensitivity study. The total and statistical uncertainties are obtained directly from the fit, while the non-linearity, the reactor spectrum shape and the difference between both likelihood implementations are added in quadrature. The largest systematic contribution comes from the background components (3.5\%), followed by the trigger efficiency (2.8\%), while the quenching factor, the non-linearity, the active mass, and the energy calibration contribute between 1 and 2\% each. The difference between Likelihood A and B contributes only 0.6\%. Compared to the expectation, the contribution of the trigger efficiency is larger, while that of the energy calibration is less than half of the expected value. The individual contributions also differ between both likelihood implementations, while the total systematic uncertainties agree. Since the nuisance parameters are correlated, their individual contributions are not independent, and the breakdown depends on the parametrization of the likelihood. The total uncertainty on $N_s$ is not affected. The random coincidences and the efficiency loss at high energy are negligible.

\begin{table*}[t!]
    \centering
    \caption{Summary of the consistency checks performed on the final \CEvNS~result. For each test, the number of signal counts $N_s$ (or the signal rate $R_s$ for the reactor-on minus reactor-off fit) is compared to the final combined fit or to the SM expectation $N_{\mathrm{SM}}$ for the corresponding configuration. All uncertainties include statistical and systematic contributions. The previous background model uses DL and TL thicknesses of 0.60\,mm and 0.90\,mm for C6 and C7 (Sec.~\ref{sec:likelihood}).}
    \label{tab:crosschecks}
    \setlength{\tabcolsep}{5pt}
    \renewcommand{\arraystretch}{1.5}
    \begin{tabular}{lcc}
        \hline\hline
        Test & Result & Assessment \\
        \hline
        Reactor-on only fit & $N_s = 1568 \pm 153$ & Agrees with final fit result \\
        Reactor-off fit with free signal & $N_s = 23 \pm 30$ & Consistent with zero \\
        Thresholds $+10$\,eV$_{ee}$ & $N_s = 1259 \pm 145$ & Consistent with SM ($N_{\mathrm{SM}} = 1369 \pm 163$) \\
        Thresholds $+20$\,eV$_{ee}$ & $N_s = 1064 \pm 143$ & Consistent with SM ($N_{\mathrm{SM}} = 1092 \pm 131$) \\
        Fit range to 2\,keV$_{ee}$ & $N_s = 1651 \pm 142$ & Agrees with final fit result \\
        Fit range to 8\,keV$_{ee}$ & $N_s = 1698 \pm 141$ & Agrees with final fit result \\
        Statistical uncertainties scaled for goodness of fit & $N_s = 1565 \pm 170$ & Agrees with final fit, $N_s/\sigma(N_s) = 9.2$ \\
        Reactor-on minus reactor-off fit & $N_s = 1318 \pm 403$ & Agrees with final fit result \\
        Previous background model (C6, C7) & $N_s = 1555 \pm 152$ & Agrees with final fit result \\
        \hline\hline
    \end{tabular}
\end{table*}

Several consistency checks were performed on the final dataset to assess the robustness of the \CEvNS~result. First, the fit was repeated using only the reactor-on spectra, dropping the likelihood contribution of the reactor-off data and model. Next, the reactor-off spectra were fitted with a free signal component, to test the validity of the background model and to exclude the misidentification of a background source as signal. Furthermore, the fit was repeated with the energy thresholds increased by 10~eV$_{ee}$ and 20~eV$_{ee}$, and with extended fit ranges up to 2~keV$_{ee}$ and 8~keV$_{ee}$. The fit was also repeated with the previous background model, using DL and TL thicknesses of 0.60~mm and 0.90~mm for C6 and C7 (Sec.~\ref{sec:likelihood}). To assess the impact of the imperfect description of the background, the statistical variance of all bins was scaled by a common factor of 1.3, corresponding to a 14\% increase of the statistical uncertainty, yielding a $p$-value above 0.1 for the combined fit. Lastly, the difference between the reactor-on and reactor-off spectra was fitted directly, with the signal rate $R_s$ in counts per day as free parameter. The results of all consistency checks are summarized in Table~\ref{tab:crosschecks}. All results are consistent with the final fit. The reactor-off fit gives a signal compatible with zero, and the ratio to the SM expectation remains stable when increasing the energy thresholds, with $N_s/N_{\mathrm{SM}} = 0.92$ and 0.97 for an increase of 10~eV$_{ee}$ and 20~eV$_{ee}$, respectively, compared to 0.92 for the default thresholds. The fits with extended ranges yield slightly higher values, by about 5--9\%, within the uncertainties. The previous background model yields $N_s = 1555 \pm 152$, in agreement with the final result. The smallest scaling of the statistical variance yielding $p > 0.1$, scanned in steps of 0.05, is a factor of 1.3, equivalent to a 14\% increase of the statistical uncertainty. It gives $\chi^2/\mathrm{ndf} = 1.07$ and a $p$-value of 0.18. With this scaling, the total uncertainty on $N_s$ increases from 155 to 170 counts, reducing $N_s/\sigma(N_s)$ from 10.1 to 9.2, while $N_s$ remains unchanged. The background-shape mismatch therefore affects the precision of the result only moderately. As a test largely independent of the background model, the reactor-off spectra, scaled to the reactor-on live time, were subtracted from the reactor-on spectra, and the difference was fitted with the predicted \CEvNS~spectral shape only, with the signal rate $R_s$ as free parameter and its distribution among the detectors fixed by the prediction. The difference in overburden between reactor-on and reactor-off periods is corrected for using the simulated muon- and neutron-induced spectra. The fit yields $N_s = 1318 \pm 403$ for the full reactor-on exposure and a significance of 3.3~$\sigma$. This agrees with the final result within 0.7~$\sigma$ and with the SM expectation within 0.9~$\sigma$. Since the background cancels in the difference, this test confirms the presence of a reactor-correlated signal at the expected rate without relying on the simulated background shape, and is therefore insensitive to the imperfect description of the background discussed above. Its larger uncertainty is driven by the limited reactor-off exposure of 86~kg$\cdot$days.

\section{\label{sec:conclusions} Conclusions and outlook}

This paper has presented the \conusplus~run 2 dataset, collected with the new detector array following the November 2024 detector upgrade described in Sec.~\ref{sec:exp}. The new 2.42~kg detectors achieve a background level per unit mass reduced by approximately a factor of two compared to the original detectors, while maintaining low energy thresholds between 140 and 170~eV$_{ee}$ and an energy-scale uncertainty of at most 3~eV$_{ee}$, thanks to the larger calibration dataset described in Sec.~\ref{sec:det}. After the refined set of waveform-based selection cuts described in Sec.~\ref{sec:analysis}, the exposure amounts to 1296~kg$\cdot$days reactor-on and 86~kg$\cdot$days reactor-off. The pulse-shape analysis, although not applied as a selection cut, provides an independent cross-check of the background model. The multi-site fraction agrees with simulation, while the excess of surface pulses is consistent with surface-related contributions not included in the simplified simulation. The likelihood framework was updated to fit directly the number of signal counts, which propagates the correlations between the signal and the nuisance parameters, and the total uncertainty on the signal prediction was reduced from 16.4\% in run 1 to 11.6\% in run 2.

The blind analysis of the run 2 dataset yields $N_s = 1565 \pm 155$ signal counts, corresponding to a relative precision of 9.9\% ($N_s/\sigma(N_s) = 10.1$), in agreement with the expected sensitivity. The ratio to the SM expectation, $N_s/N_{\mathrm{SM}} = 0.92 \pm 0.14$, is consistent with the run 1 value of $1.14 \pm 0.36$~\cite{Ackermann:2025obx}. Compared to run 1, the relative uncertainty on the number of signal counts improves by almost a factor of three, from 27\% to 9.9\%. The result is robust against the choice of likelihood implementation, the energy thresholds, the fit range, and the dead- and transition-layer thicknesses of the new detectors, and the background model is validated by the reactor-off data. The comparison with other \CEvNS~measurements and the determination of the weak mixing angle from this result are presented in the accompanying publication~\cite{ConusPlusRun2PRL}.

Looking ahead, the measurement remains statistics-dominated, with a statistical uncertainty of 7.5\% and a systematic uncertainty of 6.5\%, so that additional exposure will directly improve the precision. Assuming that the systematic uncertainty remains at its current level, the statistical uncertainty becomes comparable to the systematic one at about 1700~kg$\cdot$days reactor-on. One more year of data taking at KKL would add about 1600~kg$\cdot$days reactor-on, reducing the statistical uncertainty to about 5.0\% and the total uncertainty on $N_s$ to about 8.2\%. Beyond this point, further improvement requires reducing the systematic uncertainties. The background can be further reduced by applying the pulse-shape discrimination as a selection cut, which would remove about 20\% of the background in the ROI at a signal acceptance of about 95\%, as discussed in Sec.~\ref{sec:PSD}. The main systematic contributions can also be reduced. The background-model uncertainty is dominated by the cosmic-ray neutron component~\cite{CONUS:2024vyx, Bonet:2021wjw}, which can be constrained with additional background measurement campaigns at KKL. The energy-calibration uncertainty at the threshold position can be reduced below 2~eV$_{ee}$ with neutron irradiation campaigns, as demonstrated in~\cite{SanchezGarcia:2026nvr}, which also reduces the correlated trigger-efficiency uncertainty. The active-mass uncertainty can be reduced with dedicated measurements at MPIK. In the comparison with the SM, the quenching factor and the energy calibration at threshold remain the dominant uncertainties on the signal prediction. New quenching-factor measurements in germanium are therefore planned, which would benefit the broader low-threshold germanium community and directly improve the precision achievable in future \conusplus~analyses. The constraints on physics beyond the SM previously derived from run 1 data~\cite{CONUS:2026uhz} will substantially benefit from the improved precision of run 2, marking the transition of \conusplus~from the first detection towards precision measurements.

\begin{acknowledgments}
We thank all divisions and workshops involved at
the Max-Planck-Institut für Kernphysik in Heidelberg to set up the \textsc{CONUS+} experiment, in particular T.~Apfel, M.~Reissfelder, T.~Frydlewicz, J.~Schreiner and D.~Gro\ss{}. The authors also thank Mirion Technologies (Canberra) SAS in Lingolsheim, France, for the detector upgrades and their highly professional support.  We are grateful to Kernkraftwerk Leibstadt AG for hosting and supporting the \conusplus~experiment, with special thanks to P.~Oeschger, P.~Spahr, and P.~Marsanik.
\end{acknowledgments}

\bibliography{references}
\clearpage
\appendix
\onecolumngrid

\section{\label{sec:appendix}Background model for C3 and C6}

\begin{figure}[h!]
    \centering
    \includegraphics[width=0.48\textwidth]{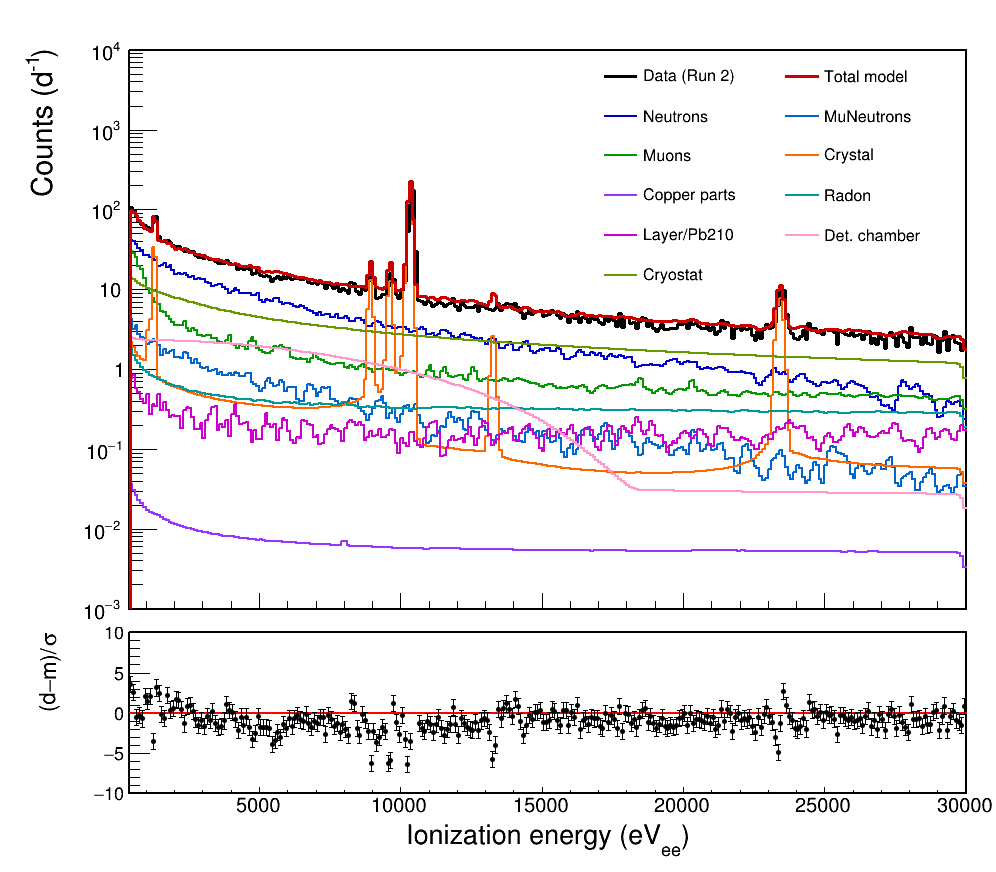}
    \hfill
    \includegraphics[width=0.48\textwidth]{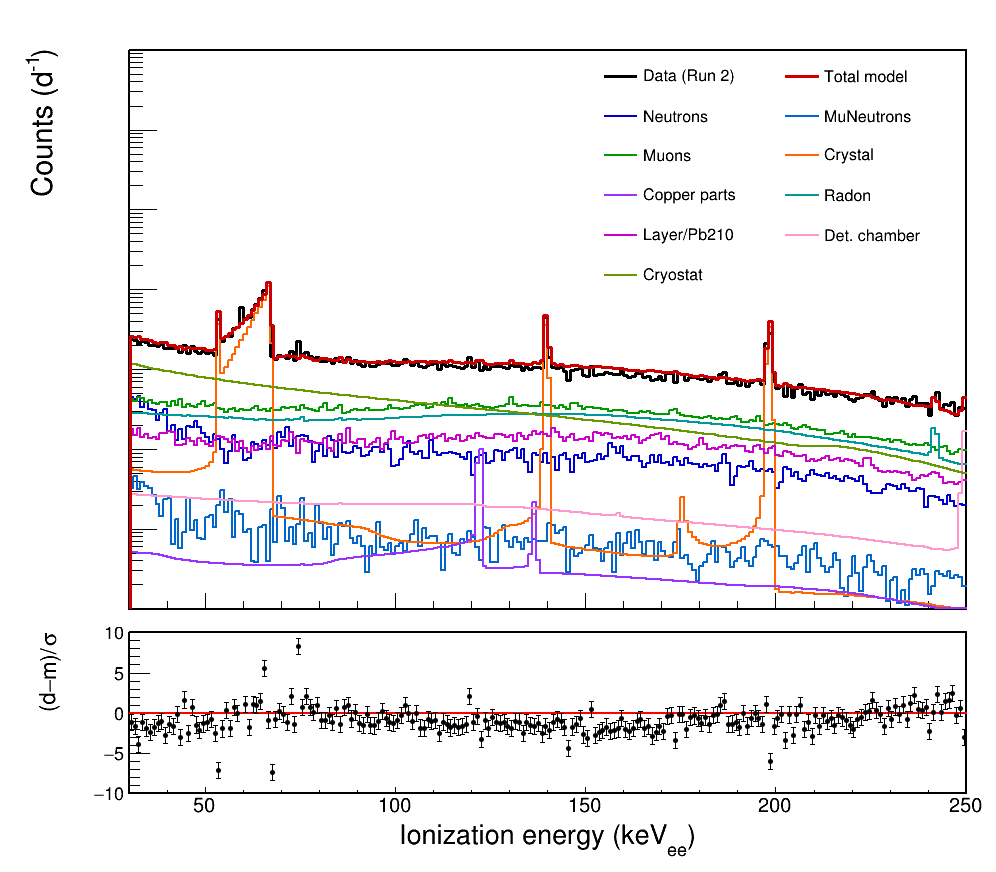}\\[2mm]
    \includegraphics[width=0.48\textwidth]{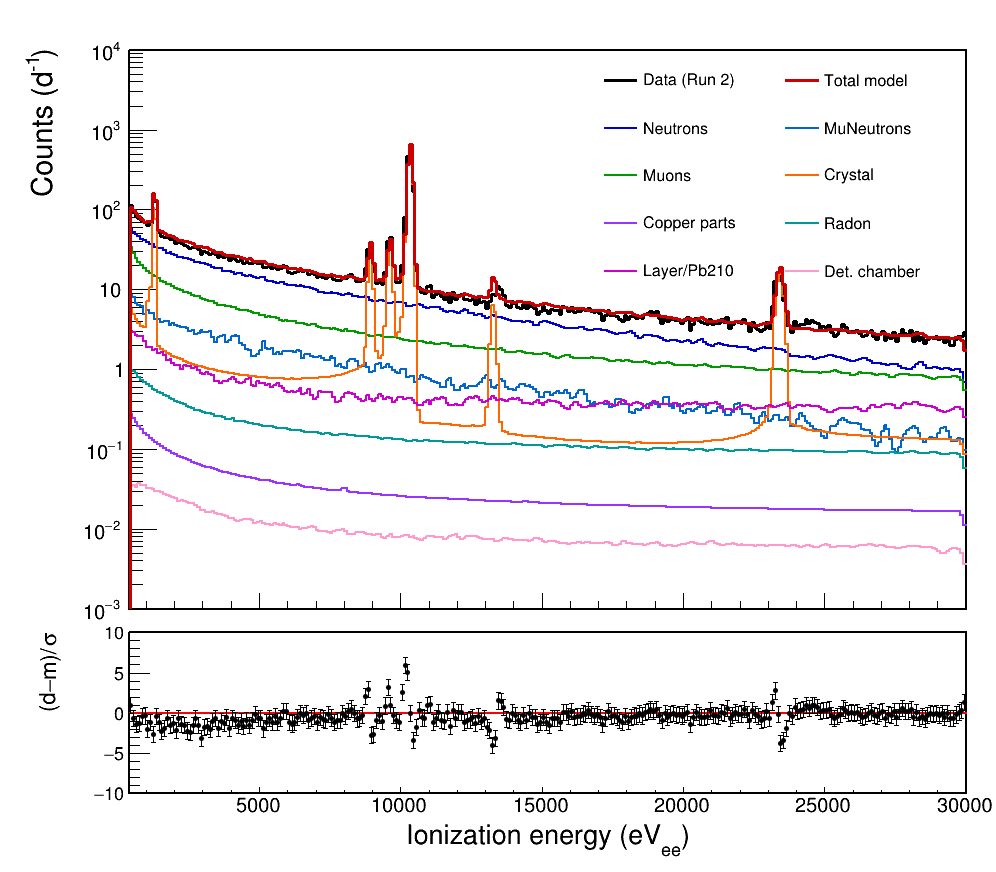}
    \hfill
    \includegraphics[width=0.48\textwidth]{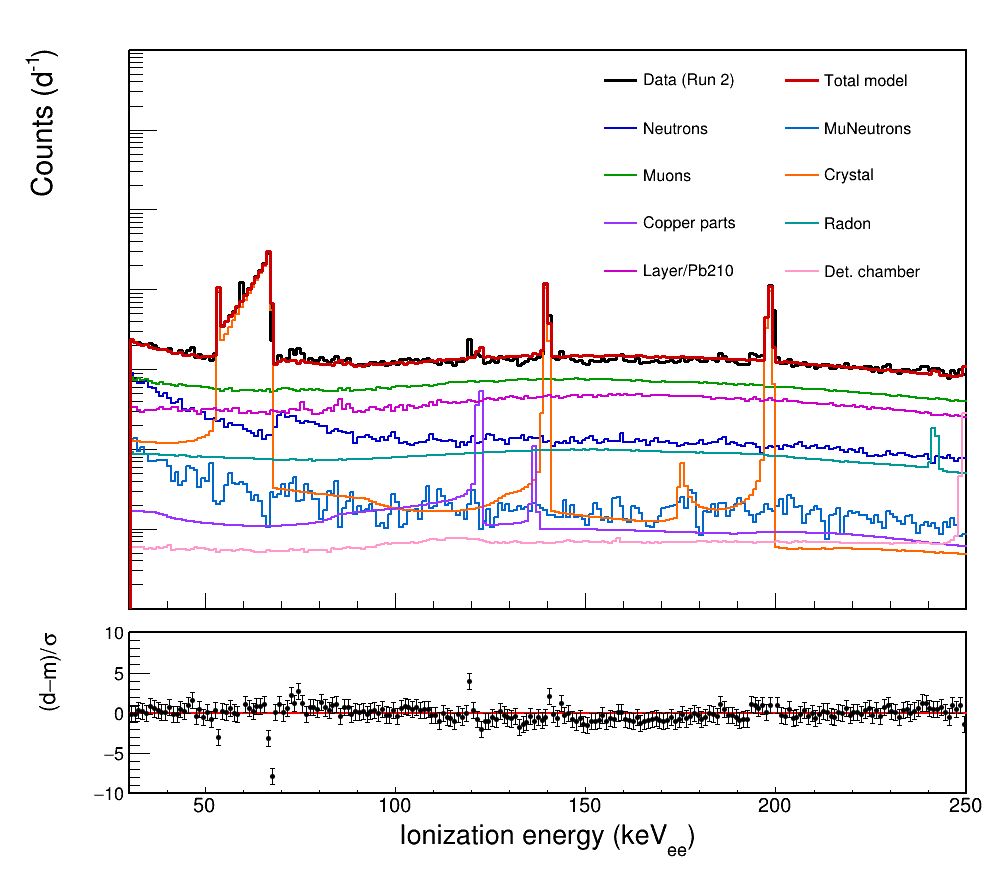}
    \caption{Background model for C3 (top) and C6 (bottom) compared to the measured energy spectrum between 0.4 and 30~keV$_{ee}$ (left) and between 30 and 250~keV$_{ee}$ (right). The individual background components are shown separately.}
    \label{fig:bkgmodel_C3C7}
\end{figure}

Figure~\ref{fig:bkgmodel_C3C7} shows the background model compared to the measured spectrum for detectors C3 and C6, complementing the C7 example shown in Sec.~\ref{sec:bkgmodel}, for both the low-energy range between 0.4 and 30~keV$_{ee}$ and the high-energy range between 30 and 250~keV$_{ee}$.

\twocolumngrid

\end{document}